\documentclass{article}
\usepackage[utf8]{inputenc}
\usepackage{comment}
\usepackage{authblk}

\usepackage{subcaption}
\usepackage{float}
\usepackage{amsmath,amssymb}
\usepackage{graphicx}
\usepackage{longtable,booktabs,array}
\usepackage{relsize}
\usepackage{calc}
\usepackage[hidelinks]{hyperref}
\title{\vspace{-3.5cm} Causal Information Exchange as a Necessary Condition for Adaptive Regulation: A Perturbational Framework for Life and Agency}
\author[1]{Brian LuValle}
\author[2]{Abicumaran Uthamacumaran}
\author[1,2,3]{Hector Zenil\thanks{Corresponding author: hector.zenil@kcl.ac.uk}}
\affil[1]{\normalsize\text{ }Algorithmic Dynamics Lab, Karolinska Institute and King's College London, Sweden and UK.}

\affil[2]{\normalsize\text{ }Oxford Immune Algorithmics, Oxford University Innovation and London Institute of Healthcare Engineering, Oxford and London, UK}

\affil[3]{\normalsize\text{ }Department of Biomedical Computing, School of Biomedical Engineering and Imaging Sciences, Faculty of Medicine and Life Sciences \& King's Institute for AI, King's College London, UK\vspace{-1cm}}

\date{}

\begin{document}
\maketitle

\begin{abstract}
Information exchange can support living organisation without providing a sufficient criterion for life. We develop a perturbational framework that distinguishes causal transmission, directional imbalance, internal retention, response structure and functional regulation. Randomised interventions define Shannon information channels; for deterministic binary recipients, their context-conditioned mutual information is exactly a paired perturbation-response indicator. Spatial and temporal averages of these local channels measure information-exposure density, rather than joint information or conserved information flux. Complementary Coding Theorem Method and Block Decomposition Method (BDM) estimates describe response and trajectory structure. A same-shaped zero-mask reference separates raw BDM from a structural contrast; neither is a channel mutual information. Retained exploratory BDM plots are distinguished from the normalised causal-channel results. Retention-weighted coupling and directionality obey algebraic bounds, and ensemble reciprocity need not imply reciprocity within individual contexts. For a specified binary sensor--internal-channel--actuator task, we derive a positive information lower bound from successful regulation and construct controllers with identical information profiles but opposite functional outcomes. A study of 2,000 ordered elementary-cellular-automaton rule pairs, six detailed examples, recurrence and lossless coding analyses, and a signed-network extension demonstrates dissociations among perturbational responsiveness, structural regularity and recurrence. The results establish causal information as a necessary resource for the stated form of adaptive regulation, while showing why neither causal coupling nor response complexity alone measures agency or life. We conclude with an operational programme for testing persistent organisation, selective reciprocal interaction and viability, explicitly separating demonstrated results from proposed biological extensions.

\noindent \textbf{Keywords:} agency; artificial life; adaptive regulation; causal intervention; mutual information; Algorithmic Information Dynamics; cellular automata.
\end{abstract}

\section{Introduction}\label{introduction}

%%%Abi: This part of the intro is jumping to low and high...it is not connecting to the initial ideas through information theory. I will work on this- there are unoperationalized definitions like "lightly compressible" ...how do the first few ideas connect to information theory. %%%

%%%what does effective compression look like and what are the standards or operationalzied definitions of compressible mechanisms- then connect to low and high information content %%%

An informational account of life must explain more than the presence of
complex patterns. A system can display statistical dependence,
persistent dynamics and sensitivity to perturbations without maintaining
itself, regulating a relevant variable or acting beneficially on its
surroundings. Conversely, effective regulation can rely on simple and
highly compressible mechanisms. The relevant distinction is therefore
not simply between low and high information content, but between
information that is present, information that can be transmitted through
causal mechanisms, and information whose use contributes to a specified
function.

A natural starting point is Schr\"odinger's negentropy argument that living systems
maintain organised states despite the tendency toward disorder
\cite{schrodinger1944}, together with Prigogine's account of dissipative
structures maintained far from equilibrium \cite{prigogine1980}. Shannon
provided the mathematical theory of communication \cite{shannon1948},
while Eigen connected information, replication and molecular evolution
\cite{eigen1971}. Statistical-physics approaches subsequently related
information flow, replication and thermodynamic constraints
\cite{england2013,parrondo2015,horowitz2014,bialek2021}. Causal-entropic
work has also explored how access to future states can generate
apparently adaptive behaviour \cite{wissnergross2013}.

The artificial-life tradition provides a complementary computational
perspective. Barricelli's early numerical evolutionary systems
\cite{barricelli1954}, Conway's Game of Life \cite{gardner1970} and
Langton's formulation of artificial life \cite{langton1989} established
that life-like organisation can be investigated in fully specified
computational worlds. Algorithmic Information Theory supplies a further
language for generative structure through Solomonoff probability and
Kolmogorov complexity \cite{solomonoff1964,livitanyi2019}. These lines of
work motivate asking not merely whether a pattern is complex, but whether
its organisation has identifiable causal consequences.

Thermodynamic and algorithmic approaches have related living
organisation to the exploitation of environmental structure rather than
to indiscriminate maximisation of randomness \cite{zenil2012}.
Information-theoretic accounts of individuality ask how a system
maintains predictive autonomy \cite{krakauer2020}, while
semantic-information approaches assess which correlations are causally
necessary for continued viability \cite{kolchinsky2018}. These
perspectives motivate an important separation: a causal communication
channel is a structural resource, whereas its contribution to survival,
regulation or agency is a further question.

A useful counterexample to attempts to define life directly from a scalar complexity measure is provided by Assembly Theory. Assembly Theory has been proposed as a way to identify biosignatures, selection and life-like
organisation from the complexity and multiplicity of assembled objects.
However, subsequent analyses have shown that its central quantities
largely reduce to established statistical compression principles, do not uniquely distinguish biological from abiotic structure, and do not by
themselves establish selection, causality or viability
\cite{uthamacumaran2024,abrahao2024,ozelim2026}. This illustrates the
central distinction pursued here: structural complexity, copy number or
compressibility may characterise an object, but life requires additional evidence about the causal organisation that maintains, regulates and couples that object to its environment.

The perturbational programme developed in earlier work provides a direct
methodological foundation for this distinction. Early computational
implementations used compressibility to characterise cellular-automaton dynamics \cite{zenil2007compressibility} and later examined their
sensitivity to controlled changes in initial conditions
\cite{zenilvillarreal2014}. Interacting cellular automata were subsequently
studied under global rules, providing an early framework for analysing how distinct dynamical subsystems influence one another
\cite{adamszenilreyesjoosten2015}. This perspective was developed further
in studies of programmability, persistent and colliding structures in
Conway's Game of Life, and systematic Algorithmic Information Dynamics
analyses of one- and two-dimensional cellular automata
\cite{zenil2015gol,zenil2019gol,zeniladams2022}.

A particularly important conceptual development is the
mutual-perturbation formulation of systems and observers introduced by
Abrah{\~a}o and Zenil \cite{abrahao2022observers}. In that framework,
observation is itself an interaction between dynamical systems, and the
appearance of emergent algorithmic information can depend on the
observer's formal knowledge while becoming robust under broader classes of description. The present work adopts this perspective by treating the
candidate agent and environment initially as observer-specified
interacting subsystems. The basic object is therefore not an isolated
agent but a pair of directed perturbational relations,

\[
A \rightleftarrows E,
\]

whose asymmetry, reciprocity, persistence and functional consequences can
be measured.

Following this mutual-perturbation
framework, we treat the
agent--environment partition as part of the observational description and identify robustness across plausible boundaries as an important future
criterion for causal individuality.

Closely related in conceptual and methodological scope are recent algorithmic approaches to persistence, regulation and agency developed
by Ruffini. His recent Game of Life study examines persistence,
regulation and the disappearance of organised patterns, while his algorithmic-regulator framework describes regulation in terms of
algorithmic complexity and model-based interaction between a regulator
and its environment
\cite{ruffini2026patterns,ruffini2026regulator,ruffini2017}. These
developments are closely aligned with the broader AIT/AID programme, but
the emphasis of the present study is different: perturbational causal information exchange across an explicit agent--environment boundary is
made the primary object of analysis, and the same intervention logic is
evaluated using both classical and algorithmic information-theoretic
tools.

There is also a methodological distinction between prediction and
intervention. Mutual information is symmetric and can be positive
because of common input. Transfer entropy conditions on the recipient's
history and quantifies additional predictive information
\cite{schreiber2000}, but predictive transfer and causal effect are not
generally interchangeable \cite{lizier2010}. Intervention-based
information flow instead asks how recipient distributions change when a source is externally varied \cite{ay2008}. Empowerment addresses the
capacity of an agent's actuation channel and therefore concerns possible
control, not necessarily beneficial realised behaviour
\cite{klyubin2005}. We build on these distinctions rather than
identifying every statistically dependent subsystem as an agent.

Algorithmic Information Dynamics combines perturbations with the
analysis of candidate generative mechanisms and their information
content \cite{zenil2019,aid2020}. AID is intervention-centred rather than
committed to a single numerical estimator. We distinguish two uses of
perturbational data. The first uses
classical Shannon information to quantify how reliably an intervention is
transmitted across a boundary. The second uses algorithmic-information
approximations: the Coding Theorem Method (CTM) estimates the algorithmic
complexity of short blocks from algorithmic probability \cite{ctm2014},
and the Block Decomposition Method (BDM) combines those local estimates
to evaluate larger objects \cite{bdm2018}. The two approaches answer
different questions. Shannon information measures an intervention channel,
while CTM/BDM characterise descriptional structure under a specified
representation. Raw BDM of an all-zero response mask is generally positive,
so a matched zero-mask reference is required for the structural contrast;
causal presence is checked separately from the mask itself. The normalised channel experiments, the
network mask descriptors and the retained exploratory ECA panels are
identified separately. Agreement between estimators is not assumed from
their shared use of perturbations.

The present study extends perturbational cellular-automaton analysis to an
explicit agent--environment partition. We ask how interventions separate
outward, inward and internally retained responses; what follows from
combining them; and what makes those quantities relevant to adaptive
regulation. Hybrid elementary cellular automata provide known local
mechanisms \cite{wolfram1984}, while a separate binary regulatory task
supplies an explicit success criterion that visual complexity cannot.

Our claim is conditional: successful challenge-dependent regulation needs
informative pathways when challenges are unpredictable, require different
responses and have no alternative causal route to the output. It does not
require positive outward dominance or a high instantaneous score in every
living state. The contribution is a protocol for measuring causal
organisation and structure, and identifying the assumptions needed for
functional interpretation.

There are three levels of inference. The channel identities and the
task-specific necessity result are mathematical statements. The numerical
experiments characterise responsiveness, structural regularity and recurrence
under stated protocols. Organisational persistence, selective reciprocity
and viability are proposed extensions that require additional observations.
In particular, a positive reciprocal score constructed from ensemble means
does not establish reciprocal interaction within the same context.

\section{An intervention-based causal information
profile}\label{an-intervention-based-causal-information-profile}

\subsection{System, boundary and intervention
ensemble}\label{system-boundary-and-intervention-ensemble}

Following the mutual-perturbation perspective of Abrah{\~a}o and Zenil
\cite{abrahao2022observers}, we regard the agent and environment not as
intrinsically privileged ontological categories, but initially as two
observer-specified interacting dynamical subsystems. Their interaction
can therefore be represented as a pair of directed perturbational
relations,

\[
A \rightleftarrows E.
\]

The quantities introduced below operationalise these two directions.
Agency is consequently not identified with interaction itself, but with
properties of the asymmetry, persistence, selectivity and functional
organisation of this mutual perturbational relation. This framing also
connects directly to earlier open-ended-evolution work in which a finite
subsystem can exhibit innovation only relative to counterfactual
isolated dynamics and interaction with an environment
\cite{adams2017,hernandez2018}.

Let a dynamical system have state \(Z_t=(A_t,E_t)\), where \(A\) is a
designated focal subsystem and \(E\) its complement. The dynamics are

\[
Z_{t+1}=F(Z_t,\xi_t). \qquad\text{(1)}
\]

The variable \(\xi_t\) represents exogenous noise when present. The
boundary, observation scale and intervention horizon are parts of the
experimental specification. They are not inferred merely by calling
\(A\) an agent.

A context \(c\) contains a pre-intervention state, the update mechanism
and the experimental settings. For a binary source cell \(i\), draw
\(U\sim\mathrm{Bernoulli}(1/2)\) independently of the context and
intervene at time \(t\) by setting

\[
\operatorname{do}(Z_{i,t}=z_i(c)\oplus U). \qquad\text{(2)}
\]

The two arms are an unchanged state (\(U=0\)) and a single-bit flip
(\(U=1\)). For a recipient observable \(R\), the experimental
distribution is

\[
p_c(u,r)=q(u)\,p\!\left(r\mid\operatorname{do}(Z_{i,t}=z_i(c)\oplus u),c\right).
\qquad\text{(3)}
\]

Define the context-specific channel score

\[
g_{i\to R}(c)=I_{p_c}(U;R),\qquad
\overline g_{i\to R}=\mathbb E_c[g_{i\to R}(c)].
\qquad\text{(4)}
\]

All logarithms are base two. Randomisation supplies the causal
interpretation: differences between the two response laws arise from the
specified intervention, subject to consistency of the simulator and
isolation of other experimental changes. The score measures the
distinguishability of experimentally induced response distributions, not
an observational association between \(A\) and \(E\). Equivalently, it
is the Jensen-Shannon divergence between the two intervention-arm
response laws:
\(g=\tfrac12 D_{\mathrm{KL}}(p^0\Vert m)+\tfrac12 D_{\mathrm{KL}}(p^1\Vert m)\),
where \(m=(p^0+p^1)/2\). It vanishes exactly when those two laws
coincide.

Future unobserved noise is marginalised in (3), not silently included in
the conditioning context. This distinction matters: for \(R=U\oplus\xi\)
with independent fair noise, \(I(U;R)=0\), even though paired outputs
always differ if the same realised noise is held fixed. Consequently,
the deterministic estimator introduced next must not be transferred
unchanged to stochastic systems.

\subsection{Exact relation to deterministic perturbation
responses}\label{exact-relation-to-deterministic-perturbation-responses}

\textbf{Proposition 1 (binary deterministic channel identity).} For a
fixed context and a deterministic binary recipient \(R=f_c(U)\),

\[
g_{i\to R}(c)=\mathbf 1\{f_c(0)\ne f_c(1)\}. \qquad\text{(5)}
\]

\emph{Proof.} Determinism gives \(H(R\mid U,c)=0\). If both intervention
arms produce the same value, \(H(R\mid c)=0\). If they produce different
binary values, the fair input makes the output equiprobable and
\(H(R\mid c)=1\) bit. Subtracting the conditional entropy proves the
result.

Thus, after fixing a context, no empirical probability table or
finite-sample mutual-information estimator is needed for a binary
deterministic ECA. We evaluate both intervention arms exactly. Sampling
uncertainty arises from the distribution of contexts, not from
estimating the two-point channel within a context.

For a cell \(j\) at lag \(\tau\), write

\[
D_{i,j,\tau}(c)=\mathbf 1\{Z^{(1,i,c)}_{j,t+\tau}\ne Z^{(0,c)}_{j,t+\tau}\}.
\qquad\text{(6)}
\]

Equation (5) makes this damage indicator an exact local
conditional-information measurement. The relation to damage-spreading
analyses is substantive: trajectory separation has long been used to
study the sensitivity of automata~\cite{derrida1986}.
We do not claim that damage spreading itself is new. The present
construction specifies its causal experiment, recipient normalisation
and separate self-retention control.

A collection of affected cells does not contain that many independent
transmitted bits. For the complete deterministic response vector under a
single binary intervention, \(I(U;\mathbf R\mid c)\) is still at most
one bit. Local scores count the spatial and temporal availability of the
same perturbation information. Their average is therefore called an
\textbf{information-exposure density}, not a joint mutual information, a
conserved information flux or an entropy-production rate.

\subsection{Outward transmission, inward transmission and internal
retention}\label{outward-transmission-inward-transmission-and-internal-retention}

For a source \(i\) and recipient region \(B\), define the geometrically
eligible site-time set

\[
\Omega_{i\to B}(T)=\{(j,\tau):j\in B,\ 1\le\tau\le T,\ d(i,j)\le\tau\},
\qquad\text{(7)}
\]

where \(d\) is periodic lattice distance. The last condition is the
radius-one ECA light-cone bound. It depends on geometry, not on whether
a particular rule actually transmits a perturbation. With a specified
distribution over source cells, define

\[
X=\mathbb E_{c,i\in P_A}\!\left[
\frac{\sum_{(j,\tau)\in\Omega_{i\to E}(T)}D_{i,j,\tau}(c)}{|\Omega_{i\to E}(T)|}
\right], \qquad\text{(8)}
\]

\[
Y=\mathbb E_{c,i\in P_E}\!\left[
\frac{\sum_{(j,\tau)\in\Omega_{i\to A}(T)}D_{i,j,\tau}(c)}{|\Omega_{i\to A}(T)|}
\right]. \qquad\text{(9)}
\]

\(X\) measures outward exposure and \(Y\) inward exposure. Each source
receives equal weight; division occurs before averaging over sources. An
empty eligible set is assigned zero and must be reported as an
unreachable probe, not interpreted as evidence of biological inactivity.
The primary experiment has no such empty cross-boundary sets.

To distinguish internal retention from an effect that leaves the
subsystem and returns through the environment, run an additional
intervention arm. At every update, overwrite all environment cells with
their unperturbed reference values, while allowing \(A\) to evolve from
its perturbed state. Denote its damage indicator by
\(D^{\mathrm{clamp}}\). Using the late-half set

\[
\Omega^{\mathrm{late}}_{i\to A}(T)=
\{(j,\tau)\in\Omega_{i\to A}(T):\tau>\lfloor T/2\rfloor\},
\qquad\text{(10)}
\]

we define

\[
S=\mathbb E_{c,i\in P_A}\!\left[
\frac{\sum_{(j,\tau)\in\Omega^{\mathrm{late}}_{i\to A}(T)}D^{\mathrm{clamp}}_{i,j,\tau}(c)}
{|\Omega^{\mathrm{late}}_{i\to A}(T)|}
\right]. \qquad\text{(11)}
\]

This is internally mediated retention of the tested perturbation, not
self-maintenance, robustness or the ability to repair damage. A
corrective mechanism can intentionally erase a single-bit perturbation.
Such erasure is not evidence that the mechanism lacks agency. The
late-half convention tests retention over a stated finite timescale; it
is not a universal timescale of life. We retain the corresponding
open-environment score, \(S_{\mathrm{open}}\), as a diagnostic of the
clamping assumption.

Each local information value is normalised by the one-bit intervention
entropy before aggregation; its numerical value is the indicator in (5).
Thus \(X\), \(Y\), \(S\) and their aggregate scores are dimensionless,
with \(0\le X,Y,S\le1\). Opportunity normalisation prevents the direct
counting of unreachable recipient cells, but it does not remove
dependence on the boundary, source distribution, accessible states or
observation horizon.

Write \(x_c,y_c,s_c\) for the source-averaged primitive quantities within
context \(c\), so that \(X=\mathbb E_c[x_c]\), \(Y=\mathbb E_c[y_c]\) and
\(S=\mathbb E_c[s_c]\). This specifies the aggregation level before any
nonlinear product is formed. The clamped experiment preserves the
unperturbed environmental drive; \(S\) measures internally mediated
retention under that drive, not autonomy without external input.

\subsection{Coupling, directionality and reciprocal
coupling}\label{coupling-directionality-and-reciprocal-coupling}

Define three derived scores:

\[
C=S(X+Y),\qquad D=S(X-Y),\qquad B=2S\sqrt{XY}. \qquad\text{(12)}
\]

\(C\) is retention-weighted coupling, \(D\) signed directional
influence, and \(B\) an ensemble reciprocal-coupling summary. These are
descriptive products of the primitive means, not established measures of
life or adaptive agency. In particular, \(D\) compares
the two cross-boundary directions; it does not compare agent influence
with the environment's influence on itself. No \(E\to E\) term occurs.

The dimensionless directionality ratio is

\[
\alpha=\frac{X-Y}{X+Y}\quad\text{for }X+Y>0. \qquad\text{(13)}
\]

When \(X+Y=0\), \(\alpha\) is undefined and is not assigned a misleading
zero. If \(S=0\) but \(X+Y>0\), \(\alpha\) remains a description of
unweighted cross-boundary asymmetry, while \(C=D=B=0\). The ratio
\(D/C\) equals \(\alpha\) only when \(C>0\).

Positive \(D\) or \(\alpha\) indicates outward dominance under the
selected protocol. When \(S>0\), zero \(D\) indicates ensemble directional balance, not
absence of influence. Zero \(\alpha\) similarly indicates equality of
the two positive cross-boundary means. If \(S=0\), however, \(D=0\)
regardless of their imbalance. Negative values indicate inward dominance, not absence
of agency. \(B>0\) holds exactly when the three primitive ensemble means
are positive. It does not imply that they are positive in the same context.
Similarly, \(C>0\) alone does not establish bidirectional interaction.

For example, an equal mixture of contexts with
\[
(x_c,y_c,s_c)=(1,0,1)
\quad\hbox{and}\quad
(x_c,y_c,s_c)=(0,1,1)
\]
has \(B=1\), although every context has zero reciprocal score. The distinct
quantities
\[
B_{\mathrm{within}}=\mathbb E_c[2s_c\sqrt{x_cy_c}],
\qquad
\pi_{\mathrm{rec}}=\Pr_c(x_c>0,\ y_c>0,\ s_c>0)
\]
would respectively describe mean within-context reciprocity and its
prevalence. Both vanish in this example. These definitions clarify the
inference; no unreported values of them are claimed for the simulations.
The numerical tables retain \(B=2S\sqrt{XY}\), computed after averaging
primitive scores. Even \(B_{\mathrm{within}}>0\) concerns separately probed
channels within a finite context and window, not proof of an operating
feedback loop, simultaneous bidirectional action or beneficial function.

\begin{figure}[ht]
\centering
\includegraphics[width=\textwidth,height=2.5in]{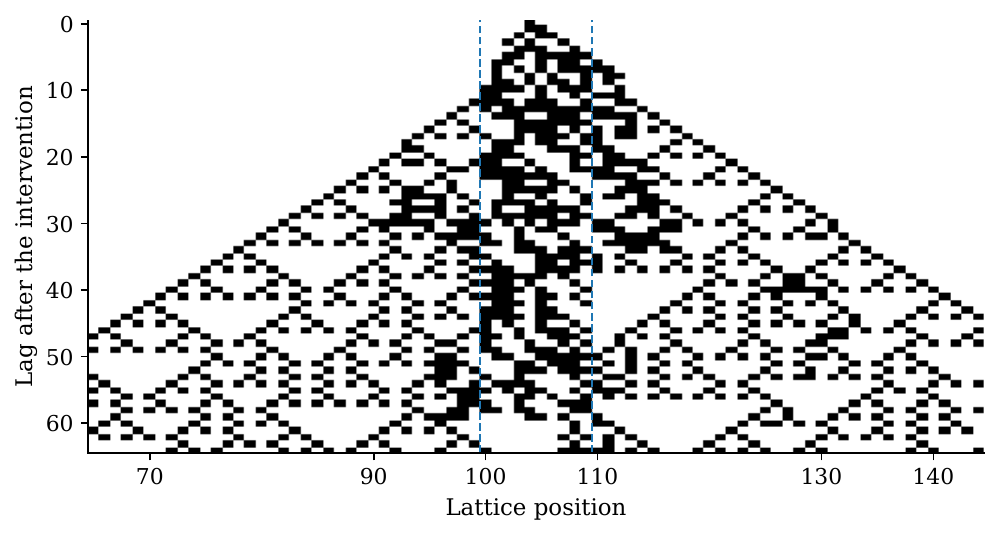}
\caption{A paired perturbation response for rule 30 in a rule-18
environment. Black cells differ from the unperturbed trajectory after a
single flip at position 104. Dashed vertical lines delimit the ten-cell
focal region; time runs downwards. This is the first stored initial
condition after 64 updates, not a selected high-scoring example. The
plotted 64-step response illustrates localisation; the primary score
uses a 32-step horizon.}
\end{figure}

\subsection{Algorithmic descriptions and the zero-mask reference}
\label{bdm-baseline}

For a finite object \(x\), Kolmogorov complexity \(K(x)\) is uncomputable.
The algorithmic coding theorem motivates CTM estimates for short objects
from output-frequency distributions of finite computational experiments
\cite{ctm2014}. BDM combines local CTM estimates with logarithmic
multiplicity terms \cite{bdm2018}. Write
\[
\mathrm{BDM}_{\theta}(x)=
\sum_{r}\bigl(\mathrm{CTM}_{\theta}(r)+\log_2 n_r\bigr),
\]
where \(r\) ranges over distinct blocks and \(\theta\) specifies the CTM
table, block dimensions, partition alignment, remainder convention and
implementation. Fixed-block BDM approximates aspects of local descriptional
structure; it does not provide an exact shortest program for the full object.

For paired recipient arrays with identical dimensions, define
\[
M_{i\to R}(c)=R^{(1,i,c)}\oplus R^{(0,c)},\qquad
K^{\mathrm{raw}}_{i\to R}(c)=\mathrm{BDM}_{\theta}(M_{i\to R}(c)).
\]
The superscript \(\mathrm{raw}\) identifies a computable descriptor rather
than exact Kolmogorov complexity. A null response has \(M_{i\to R}(c)=0\),
but its raw BDM is generally positive. The same-shaped zero-mask reference is
\[
K^0_{i\to R}(c)=
\mathrm{BDM}_{\theta}\!\left(0_{\operatorname{shape}(M_{i\to R}(c))}\right).
\]
With the same estimator settings and array convention, define
\[
\Delta K_{i\to R}(c)=
K^{\mathrm{raw}}_{i\to R}(c)-K^0_{i\to R}(c).
\]
This signed structural contrast vanishes for a null response. The converse
is not assumed: different masks can share a BDM value. The contrast is not
guaranteed nonnegative for arbitrary settings and is not a divergence,
mutual information or exact causal-effect magnitude.

Presence of a deterministic response is established separately by
\[
J_{i\to R}(c)=\mathbf 1\{M_{i\to R}(c)\ne0\}.
\]
A structural report can therefore distinguish response occurrence, exposure
density, raw BDM, the matched zero reference and the signed contrast.
Conditioning BDM summaries on \(J=1\) would further characterise non-null
responses. Neither that analysis nor new baseline-corrected values are
asserted where they have not been calculated.

The baseline also matters for directional composites. With matched array
shapes and fixed \(\theta\), source- and context-averaged outward and inward
raw descriptors share a reference \(K_0\), giving
\[
\begin{split}
K_X^{\mathrm{raw}}+K_Y^{\mathrm{raw}}
 &=\Delta K_X+\Delta K_Y+2K_0,\\
K_X^{\mathrm{raw}}-K_Y^{\mathrm{raw}}
 &=\Delta K_X-\Delta K_Y.
\end{split}
\]
The sum retains an offset; the difference cancels it only when the
references match. Unequal recipient sizes or different block conventions
invalidate this cancellation. If references differ across samples, they
must be subtracted at sample level before applying the specified weighting.
Because the contrasts can be signed, they do not inherit the
nonnegative-component bounds or square-root reciprocity formula of the
Shannon scores.

Unperturbed layer complexity, response-mask complexity and the joint
description of perturbed and reference trajectories are different objects.
In particular, \(\mathrm{BDM}(R^{(1)}\oplus R^{(0)})\) is not generally
\(K(R^{(1)})-K(R^{(0)})\) or algorithmic mutual information. A zero-array
reference for an unperturbed layer is a regularity reference, not a
no-intervention control. The retained ECA panels include layer descriptors
and native-scale composites; their role is specified separately from the
network mask analysis.

The common AID principle is controlled perturbation. A structural label
such as \emph{coupling}, \emph{life} or \emph{agency} provides neither
null calibration nor functional validation.

\subsection{Composition and collective interpretation}
Combining subsystems requires fixed update rules, source weights, recipient
sets and clamping. Exposure totals can obey accounting identities, but
normalised densities and nonlinear products are not simply additive.
Score cross-terms alone do not demonstrate emergent causation or collective
function. The complete derivation and its assumptions are retained in
Appendix~\ref{appendix-composition}.

\section{Mathematical consequences and a conditional necessity
result}\label{mathematical-consequences-and-a-conditional-necessity-result}

\subsection{Algebraic geometry is not empirical
validation}\label{algebraic-geometry-is-not-empirical-validation}

\textbf{Proposition 2 (bounds and factorisation).} For
\(X,Y,S\in[0,1]\),

\[
0\le C\le2,\quad |D|\le\min\{C,2-C\},\quad -1\le\alpha\le1,
\qquad\text{(14)}
\]

where the ratio is defined, and

\[
C^2=D^2+B^2. \qquad\text{(15)}
\]

\emph{Proof.} Non-negativity gives \(|X-Y|\le X+Y\), hence \(|D|\le C\).
Also \(C+|D|=2S\max(X,Y)\le2\), which gives the sharper bound
\(|D|\le2-C\). Expanding \((X+Y)^2-(X-Y)^2=4XY\) proves (15). The ratio
bound follows from the same non-negativity argument.

For \(C>0\), \(B=C\sqrt{1-\alpha^2}\). These are dependent summaries of
the same three components, not orthogonal empirical measurements. A
scatterplot of \(D\) against \(C\) must lie within the bounds; observing
that high absolute \(D\) requires high \(C\) therefore supplies no
independent evidence that life precedes agency.

The stronger bound also exposes a conceptual problem with equating
agency to imbalance. Maximal coupling, \(C=2\), forces \(X=Y=S=1\) and
therefore \(D=0\). A perfectly reliable, balanced controller would
receive the largest coupling score and zero directional imbalance.
Conversely, \(|D|=C>0\) implies that one cross-boundary direction is
absent and \(B=0\). Large imbalance is not a general objective for
living organisation.

\subsection{What information is genuinely necessary for
regulation?}\label{what-information-is-genuinely-necessary-for-regulation}

A useful necessity claim must specify the task rather than define
success by a high score. Consider a fair binary challenge \(W\)
independent of a background context \(K\). Let \(M\), \(Q\) and \(R\) be
binary sensor-memory, action and external response variables, with
causal chain

\[
W\longrightarrow M\longrightarrow Q\longrightarrow R
\quad\text{conditional on }K. \qquad\text{(16)}
\]

Assume independent channel noise conditional on \(K\), no direct
challenge-to-response bypass, and no unmeasured common cause between
adjacent channel variables. The action is evaluated against the
externally specified requirement \(R=W\). There is no
challenge-independent response that exceeds one-half success.

\textbf{Theorem 3 (task-conditioned causal-information necessity).} Let
\(\varepsilon=\Pr(R\ne W)<1/2\). With \(h_2\) the binary entropy
function and \(b=1-h_2(\varepsilon)\),

\[
I(W;M\mid K)\ge b,\quad I(M;Q\mid K)\ge b,\quad I(Q;R\mid K)\ge b.
\qquad\text{(17)}
\]

If adjacent-channel interventions use the task-induced input
distributions, define the task-aligned scores \(Y_*=I(W;M\mid K)\),
\(S_*=I(M;Q\mid K)\) and \(X_*=I(Q;R\mid K)\). Information values are
expressed relative to the same one-bit reference. They then obey

\[
C_*=S_*(X_*+Y_*)\ge2b^2>0,\qquad
B_*=2S_*\sqrt{X_*Y_*}\ge2b^2>0. \qquad\text{(18)}
\]

\emph{Proof.} Conditional binary Fano bounds give
\(H(W\mid R,K)\le h_2(\varepsilon)\). Since \(W\) is fair and
independent of \(K\), \(I(W;R\mid K)\ge1-h_2(\varepsilon)\). Conditional
data processing along (16) bounds this end-to-end information above by
the information across each of the three cuts. The causal-chain
assumptions make the adjacent interventional channels coincide with
these conditional distributions when their input distributions match.
Applying the three lower bounds to the two products yields (18). These steps use
standard information inequalities~\cite{cover2006}.

This is a task-specific necessity result, not an algebraic restatement
of (12). Success is defined independently, and the lower bound depends
on measured error. It does not require \(D_*>0\); the benchmark below
has \(D_*=0\) while regulating successfully.

The information quantities in (17) average over \(K\), so the resulting
\(B_*>0\) is an aggregate statement, not a guarantee for every context.
For a particular \(k\) whose task error \(\varepsilon_k<1/2\), the same
proof applies within that context because \(W\) remains fair given \(K=k\).
Overall success does not imply that every context meets that error condition.

Several conditions are essential. A direct environmental pathway to the
response can invalidate the chain. Predictable challenges, pre-existing
information or a single response that succeeds under every challenge
remove the one-half open-loop bound. A different task requires a
corresponding decision-theoretic bound. Most importantly, the
task-aligned whole-channel quantities in (17) are not automatically
equal to spatially averaged single-bit ECA scores. The latter are
experimental diagnostics, not universal necessary conditions for every
possible regulator.

Here \(S_*\) measures information across the internal sensor-to-action
channel; it does not itself measure temporal memory duration or
organisational survival. Connecting the theorem to an ECA would require an
explicit task implementation, task-induced intervention distributions and
recipient representations that retain the relevant information.

\subsection{An explicit failure of
sufficiency}\label{an-explicit-failure-of-sufficiency}

\textbf{Proposition 4 (identical information, opposite function).} There
are controllers with identical \((X_*,Y_*,S_*,C_*,D_*,B_*,\alpha_*)\)
but opposite performance on the same regulatory task.

\emph{Proof by construction.} Let \(W\) be a fair bit and introduce
independent sensor, memory and actuator errors \(N_s,N_m,N_a\) with
probabilities \(e_s,e_m,e_a\). Define

\[
M=W\oplus N_s,\qquad Q=M\oplus N_m\oplus p,\qquad R=Q\oplus N_a,
\qquad\text{(19)}
\]

where policy \(p=0\) is corrective and \(p=1\) inverts the action. In
both policies,

\[
Y_*=1-h_2(e_s),\quad S_*=1-h_2(e_m),\quad X_*=1-h_2(e_a).
\qquad\text{(20)}
\]

The policy flip is bijective and cannot change these information
quantities. However, with

\[
q=\frac{1-(1-2e_s)(1-2e_m)(1-2e_a)}{2}, \qquad\text{(21)}
\]

task success is \(1-q\) for the corrective policy and \(q\) for the
inverting policy. At zero noise, both have \(C_*=B_*=2\) and \(D_*=0\),
while success is respectively one and zero. No function of this
information profile alone can distinguish their functional correctness.

Response-mask structure does not automatically repair this ambiguity.
Complementing corresponding binary outputs in both intervention arms
leaves their XOR mask unchanged:
\[
(\neg R^{(1)})\oplus(\neg R^{(0)})=R^{(1)}\oplus R^{(0)}.
\]
Both its raw BDM and matched zero-mask contrast are therefore unchanged by
that relabelling, although task correctness can reverse. This statement
concerns matched paired masks; it is not an assertion that every possible
algorithmic descriptor is invariant to every policy change.

The counterexample does not make causal information irrelevant. It
identifies what is missing: a relation between the causal mechanism and
an independently specified regulatory requirement. A biological
extension must similarly test the effect of causal pathways on viability
rather than rename their existence as life. This is consistent with, but
not equivalent to, the viability-intervention approach to semantic
information~\cite{kolchinsky2018}.

\subsection{Boundary dependence and
non-additivity}\label{boundary-dependence-and-non-additivity}

With matched source and recipient protocols on equal-sized complementary
regions, exchanging the two regions exchanges \(X\) and \(Y\), so
\(\alpha\) changes sign. The focal retention term generally changes too;
\(D\) need not simply negate when \(S\) is recomputed for the new focal
subsystem. Padding the lattice with cells outside the relevant causal
ancestry cannot change a score, provided both the perturbation cones and
their required initial-state ancestry are preserved. This is a locality
property, not unrestricted size invariance.

Subagent scores are not generally additive. If two subagents act on the
same recipient, adding exposure counts can double-count that recipient.
Under a simultaneous intervention ensemble, let independent fair inputs
\(U_1,U_2\) produce \(R=U_1\oplus U_2\). Then \(I(U_1;R)=I(U_2;R)=0\)
but \(I((U_1,U_2);R)=1\). Joint interventions can therefore reveal
synergistic influence absent from marginal channels. Single-source
interventions that condition on the other input constitute a different
experiment and correctly give a different answer.

\section{Computational methods}\label{computational-methods}

\subsection{Experimental design}
The primary experiment uses a periodic 210-cell lattice: ten focal cells
and 200 environmental cells. Each region follows its assigned ECA rule,
with synchronous updates and actual cross-boundary neighbour states.
For rule number \(r_j\),
\[
z_j(t+1)=\left\lfloor\frac{r_j}{2^{4z_{j-1}(t)+2z_j(t)+z_{j+1}(t)}}\right\rfloor\bmod 2,
\qquad\text{(22)}
\]
The scan samples 2,000 distinct ordered rule pairs uniformly without
replacement, with 16 independent Bernoulli-one-half initial configurations
per pair. Interventions occur after 64 updates, probing every focal cell
and the five environmental cells adjoining each interface. The horizon
is 32 updates; clamped retention uses lags 17--32.

Six detailed pairs, \((30,18),(250,30),(30,2),(30,45),(110,60),(45,2)\),
are followed from 128 initial configurations at ages 0, 16, 64, 128, 256
and 512. The same configurations are followed across ages. Homogeneous
rules, sham and boundary-cut interventions, geometry changes, role
exchange, locality-preserving padding and a scalar implementation provide
controls. Full settings and validation procedures are in
Appendix~\ref{appendix-methods}; the study was not preregistered.

\subsection{Structural and network measurements}
Site-frequency entropy and level-nine raw DEFLATE describe fixed
64-update unperturbed ECA windows. Complete-state recurrence is sought
through update 4,096; undetected recurrence is censored.
Neither code length nor recurrence is a channel-information estimate.

The separate network experiment uses 32 binary nodes on a degree-four
ring, split into equal regions, with symmetric edge weights in
\(\{-1,+1\}\). Five positive-edge probabilities from zero to one each
have 64 independent network--initial-state contexts. Threshold updates
retain the previous state at ties. Boundary probes begin at age 16,
with a 16-update horizon and clamped retention at lags 9--16.
The full update and sampling protocol is in
Appendix~\ref{network-methods}.

Network BDM describes full \(16\times16\) XOR response masks. Its reported
\(K_X+K_Y\) is raw; the matched structural contrast would subtract
\(2K_0\), with \(K_0=\mathrm{BDM}_{\theta}(0_{16\times16})\).
No numerical baseline-corrected values are claimed. Estimator settings
and shape must match when using this reference. Exploratory ECA panels
use different native encodings and are retained separately in
Appendix~\ref{exploratory-results}.

\subsection{Regulation and uncertainty}
The binary task uses sensor, internal-channel and actuator errors
\((0.10,0.05,0.10)\), plus seven equal-error settings from zero to
one-half, with 100,000 challenges per policy and setting.
Corrective and inverting policies share challenges and noise; an
independent fair-action control succeeds at chance. Wilson intervals
describe individual success probabilities.

Primitive context scores are averaged before forming \(C,D,B,\alpha\).
Thus \(B\) is an ensemble summary, not within-context reciprocity.
Intervals use 5,000 bootstrap resamples of complete initial-condition
units, or complete network--initial-state units; cells, lags and probes
are not independent replicates. Temporal contrasts are paired across
ages. Ordinary intervals are 95\%; six designated endpoint contrasts use
99.167\% marginal intervals for nominal Bonferroni familywise coverage
of 95\%, subject to bootstrap approximation. Sign counts and split-half
ranks describe the sampled ensembles. Full statistical conventions,
seeds and software details are in Appendices~\ref{appendix-methods}
and~\ref{appendix-a.-estimation-algorithm-and-audit-trail}.

\section{Results}\label{results}

\subsection{The rule-pair scan separates coupling from directional
dominance}\label{the-rule-pair-scan-separates-coupling-from-directional-dominance}

The 2,000-pair scan contains 1,625 positive-coupling pair means and 375
zero-coupling means. Positive signed directionality occurs in 725 pairs
(36.25\%), negative directionality in 900 (45.00\%), and zero in 375
(18.75\%). Median coupling is 0.00390 and mean coupling is 0.02261.
There is therefore no predominance of positive directional influence in
this protocol. Unequal compartment geometry and the chosen intervention
ensemble remain part of the interpretation; the sign distribution is not
evidence for a membrane-like mechanism.

All points satisfy \(|D|\le C\) and the factorisation \(C^2=D^2+B^2\) to
numerical precision. These observations verify the implementation of the
definitions, not the biological meaning of the scores. Split-half
Spearman correlations are 0.975 for \(C\) across all pairs and 0.952 for
\(\alpha\) across the 1,956 pairs with the ratio defined in both halves.
The high repeatability supports stability of these ensemble summaries
within the sampled protocol; it does not remove bias from an
inappropriate boundary or task definition.

\begin{figure}[ht]
\centering
\includegraphics[width=4.5in,height=3in]{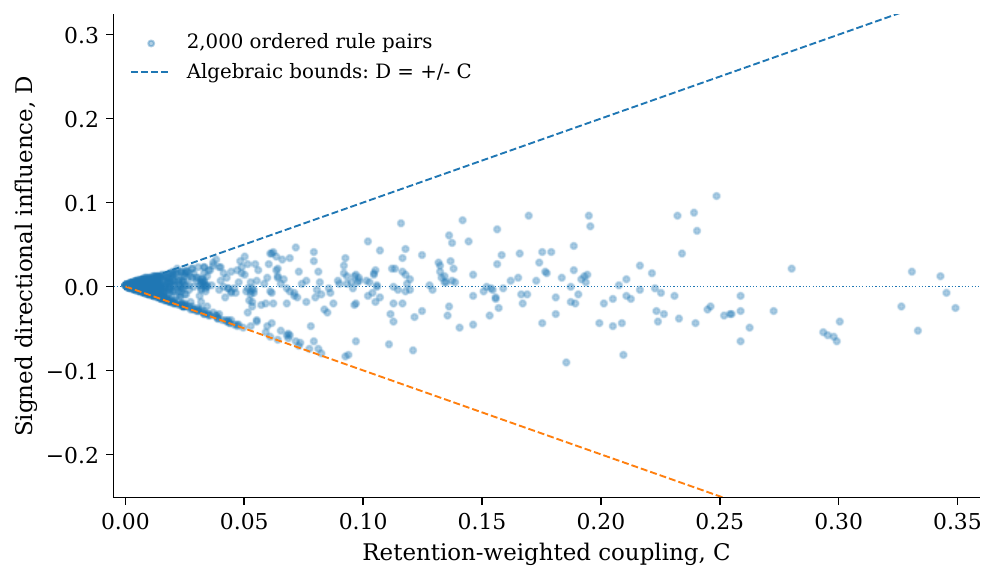}
\caption{Retention-weighted coupling and signed directional influence
for 2,000 distinct ordered rule pairs. Each point is computed from the
mean primitive scores over 16 independent initial conditions after 64
updates. Dashed lines show the algebraic bounds \(D=\pm C\) in the
displayed range. Accumulation near the origin and the bounded wedge are
not independent evidence for a life-agency hierarchy.}
\end{figure}

\subsection{Detailed cases and controls}
Rule 250 in rule 30 has outward exposure \(X=0.059\) but zero inward
exposure and late retention at age 64. Consequently \(C=0\), despite
\(\alpha=1\): a large directional ratio need not signify internally
sustained organisation. Rule 30 in rule 2 instead has positive primitive
means and aggregate \(B\), without establishing reciprocity within every
context. Rule 30 in rule 45 and rule 110 in rule 60 are inward dominated.

The controls reinforce this distinction. Constant rule 0 has zero scores;
identity rule 204 retains the altered bit but has no boundary transmission;
shift rule 170 transmits without late focal retention. All have \(C=0\).
Additive rule 90 nevertheless has \(C=0.1226\) without a regulatory task.
Complete six-case estimates, intervals, trajectories and homogeneous
controls appear in Appendices~\ref{appendix-extended-results}
and~\ref{appendix-c.-complete-example-trajectories-and-additional-results}.

\subsection{Transient changes separate coupling from directionality}
Between ages zero and 512, \(C\) falls from 0.0749 to 0.0558 for rule
110/60 and from 0.0442 to 0.0147 for rule 45/2. Both decreases survive
the multiplicity adjustment, while neither signed-\(D\) change is
resolved by its ordinary 95\% interval. Rule 30/45 has no clearly
resolved endpoint change in either score; rule 30/18 increases in
coupling. There is no general monotonic age--coupling law.
A negative directional score approaching zero increases in signed value
while decreasing in magnitude. Full paired contrasts and additional
trajectories are in Appendix~\ref{transient-changes-in-coupling-do-not-imply-monotonic-changes-in-directionality}.

\begin{figure}[ht]
\centering
\includegraphics[width=4in,height=2.5in]{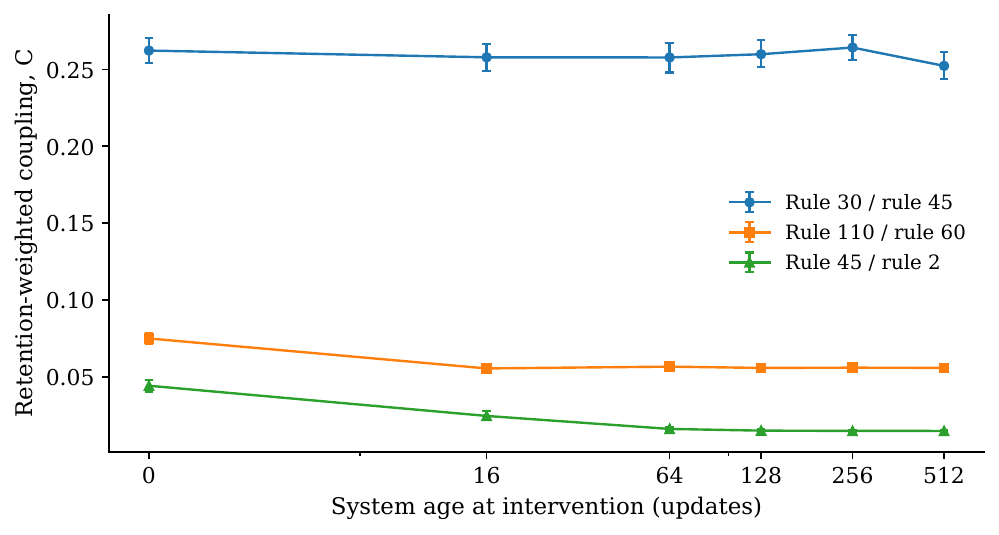}
\caption{Coupling at successive intervention ages for the three
transient comparisons. Points are calculated from 128 initial-condition
units and error bars are pointwise 95\% bootstrap intervals. The
horizontal scale is linear through age 16 and logarithmic thereafter.
Different initial conditions are not substituted at later ages.}
\end{figure}

\begin{figure}[ht]
\centering
\includegraphics[width=4in,height=2.5in]{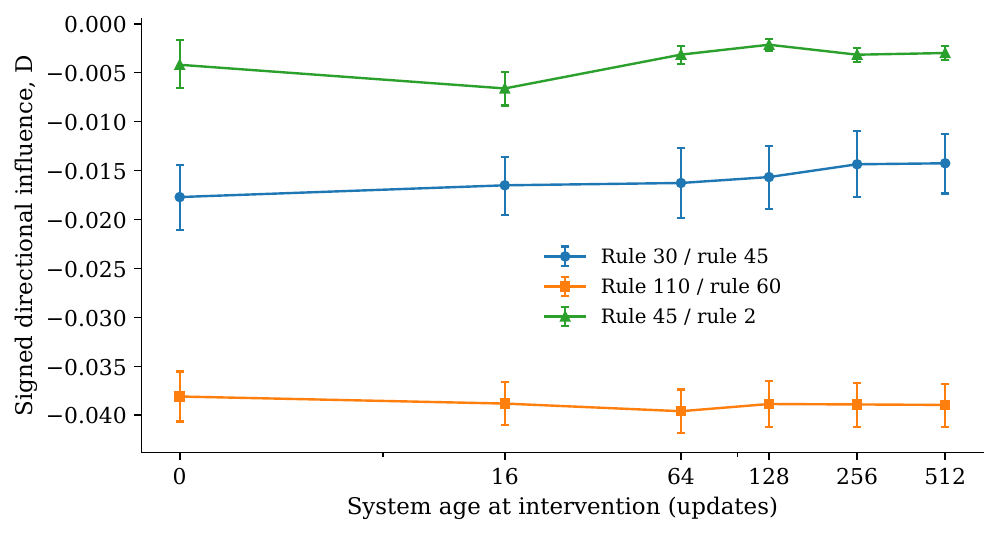}
\caption{Signed directional influence for the same initial-condition
ensembles. The rule-110/rule-60 and rule-45/rule-2 coupling decreases in
the preceding figure do not establish corresponding decreases in signed
directionality. Error bars are pointwise 95\% bootstrap intervals and do
not constitute a test of the entire time-course shape.}
\end{figure}

\subsection{Structure, recurrence and causal responsiveness}
For rule 30/2, focal site-frequency entropy remains near one bit, while
the DEFLATE ratio falls from 0.6183 to 0.1396 between ages zero and 512.
Coupling changes only from 0.0757 to 0.0727, with a paired interval
spanning zero. Structural regularisation therefore need not remove causal
responsiveness. Rule 110/60 shows the converse contrast: its code ratio
changes little while coupling decreases.

Complete-state recurrence occurs in all tested contexts for four of the
six detailed pairs. Rule 110/60 has median transient 44.5 and period
256, illustrating why short-window irregularity need not mean
nonperiodicity. The full recurrence table and interpretation are in
Appendix~\ref{exact-recurrence-and-structural-regularity-are-different-from-causal-responsiveness}.

\begin{figure}[ht]
\centering
\includegraphics[width=4in,height=2.5in]{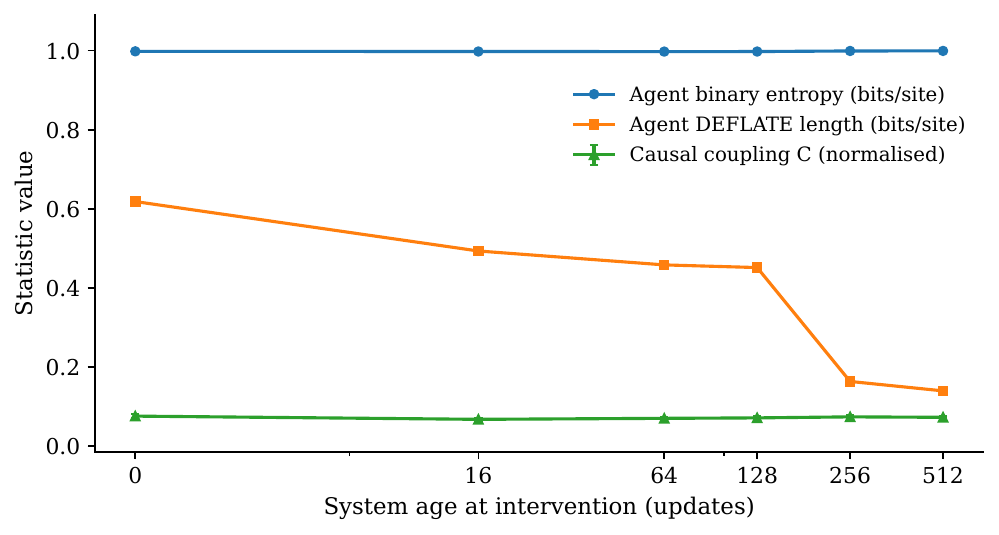}
\caption{Structural regularity and causal coupling for rule 30 in rule
2. Binary entropy, raw DEFLATE bits per original agent cell, and
normalised coupling are displayed as separately defined quantities; they
are not alternative estimators of the same object. Entropy remains close
to one, code length declines and coupling is comparatively stable. The
code and entropy use 64-update windows; the causal score uses the fixed
32-update protocol.}
\end{figure}

\subsection{Geometry and intervention controls}
Changing focal width can reverse directionality: rule 45/2 changes from
\(\alpha=-0.438\) at width six to \(+0.115\) at width twenty.
This changes the update-rule allocation as well as the boundary, so it
tests architecture dependence, not solely observational repartitioning.
Padding, matched role exchange and scalar checks support the stated
invariances. Sham and boundary-cut controls give zero Shannon effects;
a sham BDM mask instead has raw reference \(K^0\) and zero structural
contrast. Clamping may either increase or suppress a response.
Full settings, estimates and the geometry figure are in
Appendix~\ref{geometry-and-intervention-controls}.

\subsection{Equal causal information can support corrective or
harmful
responses}\label{equal-causal-information-can-support-corrective-or-harmful-responses}

\begin{longtable}[]{@{}
  >{\raggedright\arraybackslash}p{(\columnwidth - 6\tabcolsep) * \real{0.1500}}
  >{\raggedright\arraybackslash}p{(\columnwidth - 6\tabcolsep) * \real{0.1400}}
  >{\raggedright\arraybackslash}p{(\columnwidth - 6\tabcolsep) * \real{0.1900}}
  >{\raggedright\arraybackslash}p{(\columnwidth - 6\tabcolsep) * \real{0.5200}}@{}}
\caption{Regulatory benchmark at sensor, memory and actuator errors of
0.10, 0.05 and 0.10. Corrective and inverting policies have identical
causal-information profiles. The open-loop action is independent of the
memory, so its task-aligned retention channel and coupling are zero.
Each row uses 100,000 trials.}\tabularnewline
\toprule\noalign{}
\begin{minipage}[b]{\linewidth}\raggedright
Policy
\end{minipage} & \begin{minipage}[b]{\linewidth}\raggedright
\(C_*\)
\end{minipage} & \begin{minipage}[b]{\linewidth}\raggedright
Exact success
\end{minipage} & \begin{minipage}[b]{\linewidth}\raggedright
Simulated success {[}95\% interval{]}
\end{minipage} \\
\midrule\noalign{}
\endfirsthead
\toprule\noalign{}
\begin{minipage}[b]{\linewidth}\raggedright
Policy
\end{minipage} & \begin{minipage}[b]{\linewidth}\raggedright
\(C_*\)
\end{minipage} & \begin{minipage}[b]{\linewidth}\raggedright
Exact success
\end{minipage} & \begin{minipage}[b]{\linewidth}\raggedright
Simulated success {[}95\% interval{]}
\end{minipage} \\
\midrule\noalign{}
\endhead
\bottomrule\noalign{}
\endlastfoot
Corrective & 0.7579 & 0.7880 & 0.7880 {[}0.7855, 0.7905{]} \\
Inverting & 0.7579 & 0.2120 & 0.2120 {[}0.2095, 0.2145{]} \\
Open-loop & 0 & 0.5000 & 0.5009 {[}0.4978, 0.5040{]} \\
\end{longtable}

At \((e_s,e_m,e_a)=(0.10,0.05,0.10)\), both task-responsive policies have
\(X_*=Y_*=0.5310\), \(S_*=0.7136\), \(C_*=B_*=0.7579\) and
\(D_*=\alpha_*=0\). Nevertheless, their theoretical task-success
probabilities are 0.788 and 0.212. The independent-action control is at
chance. These differences are not attributable to different amounts of
channel information: the corrective and inverting systems differ only in
the policy's relation to the task requirement.

The equal-error sweep confirms the same separation across the entire
tested noise range. As channel information increases, corrective
performance improves and inverting performance worsens, despite
identical information profiles at every matched noise level. The example
directly refutes sufficiency of the task-aligned \((X_*,Y_*,S_*)\) and
their derived scores for useful regulation; it is not a direct task test
of the ECA exposure densities. It also refutes positive directional dominance as a
necessary feature of a successful regulator.

\begin{figure}[ht]
\centering
\includegraphics[width=4.5in,height=3in]{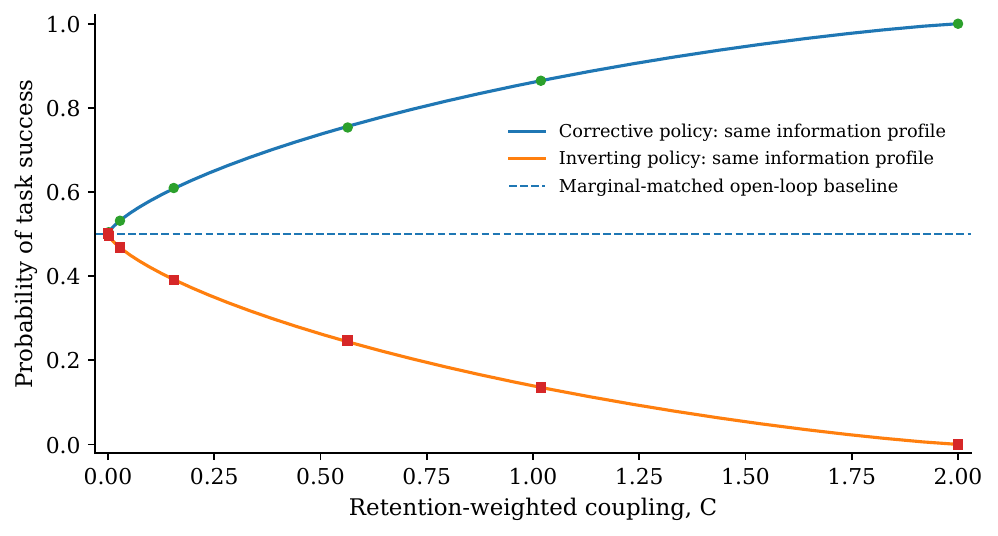}
\caption{Opposite functional outcomes at identical coupling. Curves are
the exact equal-error predictions from equations (20)--(21); points and
Wilson error bars show 100,000 simulated challenges per policy and error
setting. Corrective and inverting policies have the same information
profile at each horizontal position. The dashed line is the expected
marginal-matched open-loop success probability, shown as a chance
reference; the open-loop controller itself has zero coupling.}
\end{figure}

\subsection{Scope of the exploratory information and BDM comparisons}
Appendix~\ref{exploratory-results} retains all exploratory ECA MI-density,
BDM-composite and layer-level panels with their full interpretation.
Their native scales and incompletely matched protocols do not identify
them with the normalised channel scores. Some directional patterns
differ between MI and BDM, so the panels do not establish estimator
agreement. Their values remain raw: zero-mask references must be applied
to the encoded primitives under the actual aggregation formula, not
inferred by subtracting an arbitrary constant from a plotted product.
Layer regularity is distinct from a null causal effect, and stable
structural curves do not establish homeostasis or equilibrium.

\subsection{Signed-network proof-of-concept}
In the deterministic 32-node extension, aggregate coupling at \(p=0\)
was \(C=0.001750\), with 95\% context-bootstrap interval
\([0.001353,0.002223]\). At \(p=1\), \(C=0.000749\), with interval
\([0.000508,0.001045]\). Across the five separately sampled ensembles,
the point estimates of \(C\) decreased monotonically. This is a
descriptive ordering, not an established monotonic population law.
All aggregate \(D\) intervals included zero: the estimates do not
resolve an ensemble directional imbalance, but neither prove equality
of the channels nor demonstrate balanced interaction within contexts.

The reported BDM response-mask sum \(K_X+K_Y\) was
\(70.45\;[67.93,72.96]\) at \(p=0\) and
\(60.28\;[58.16,62.43]\) at \(p=1\). These are raw description-length
estimates, including the contribution of unchanged response masks, not
zero-referenced effect sizes. With the common \(16\times16\) shape and
fixed estimator \(\theta\), the corresponding structural contrast is
\(K_X+K_Y-2K_0\), where
\(K_0=\mathrm{BDM}_{\theta}(0_{16\times16})\).
A common subtraction preserves the between-ensemble ordering and
differences, but absolute corrected values cannot be supplied without
the numerical reference and complete estimator settings. No corrected
values or corrected intervals are claimed here. In particular, a
positive raw sum does not by itself show a nonzero causal response:
that evidence comes from the paired masks or Shannon channels.

Recurrence was observed in all 64 contexts at each mixture within the
512-update search window, with detected periods of one or two updates
(period one only at \(p=1\)). The channels and raw structural descriptors
nevertheless varied across the sign ensembles. The experiment thus
extends the specified perturbational protocol beyond cellular automata
and illustrates the non-equivalence of recurrence, channel exposure and
response description. It does not estimate within-context reciprocity,
learning, beneficial agency or a physical critical transition.

\begin{figure}[H]
\centering
\includegraphics[width=0.98\textwidth]{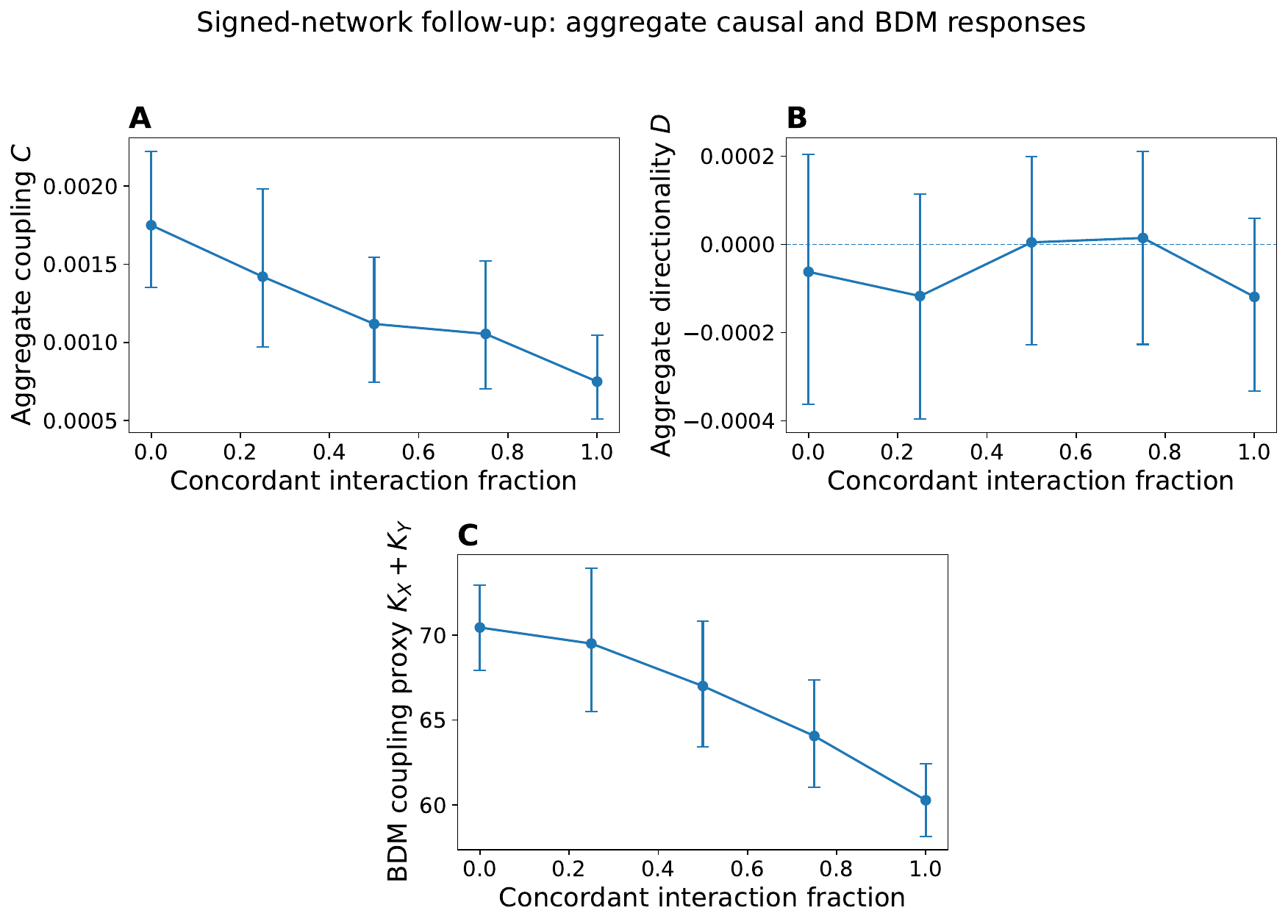}
\caption{Signed-network proof-of-concept: 32 nodes on a degree-four ring,
equal 16-node regions, symmetric random edge signs and 64 independent
network--initial-state contexts per positive-edge probability \(p\).
Panels A and B show \(C=S(X+Y)\) and \(D=S(X-Y)\), computed from
primitive ensemble means. Balance of means is not within-context reciprocity.
Panel C's BDM proxy denotes the raw mean \(K_X+K_Y\) of XOR-mask
descriptions, not causal coupling or a zero-referenced result.
With the fixed \(16\times16\) encoding, the structural contrast would
subtract \(2K_0\). Bars are 95\% percentile intervals from 5,000
complete-context bootstrap resamples, without estimator uncertainty.}
\label{fig:multiagent-aggregate}
\end{figure}

The complete five-ensemble numerical table is retained in
Appendix~\ref{appendix-network-table}.

\section{Discussion}\label{discussion}

The central result is a separation between causal access, structural
description and functional accomplishment. Theorem 3 derives positive
information across each cut of a specified challenge--sensor--internal
channel--response chain from better-than-chance regulation. Its force
comes from an independently defined task, not from naming a positive
score life or agency. Proposition 4 then supplies the complementary
limitation: controllers with identical task-aligned information profiles
can implement correct or inverted responses. Thus causal information is
necessary under the theorem's assumptions, but its amount does not
determine its use.

This necessity statement must not be transferred unchanged to the ECA
densities. The theorem concerns whole task-aligned channels under
task-induced input distributions. The ECA experiment averages local
binary response indicators over specified sources, recipients, lags and
contexts. Those averages measure causal information-exposure density,
not joint information, conserved flux or a demonstrated regulatory
capacity. The numerical work establishes properties of these diagnostic
profiles, whereas the theorem establishes a conditional constraint on
a separate explicit task. An ECA test of that constraint would first
need an implemented task and the corresponding whole-channel variables.

Aggregation also limits inference. Reported \(B=2S\sqrt{XY}\) can be
positive when different one-way contexts supply the inward and outward
means. Near-zero \(D\) can likewise reflect cancellation rather than
within-context balance. Claims about individual contexts require their
joint scores, \(B_{\mathrm{within}}\), prevalence or distributions.
Even those separately probed directions do not demonstrate operating
feedback; that requires temporal and task evidence. The present bootstrap
describes ensemble summaries, not these unmeasured properties.

BDM requires a parallel distinction. An unchanged mask has raw reference
\(\mathrm{BDM}_{\theta}(0_{\mathrm{shape}(M)})\), not necessarily zero.
Its matched contrast vanishes for a null effect but may be signed or
vanish for other masks. Causal presence is checked separately. Moreover,
complementing both intervention outputs preserves their XOR mask while
potentially reversing task correctness; structural calibration does not
resolve functional ambiguity.

The network table therefore retains its raw values. Common-baseline
subtraction preserves matched-ensemble differences, not a causal
interpretation of positive raw levels. The exploratory ECA panels use
different native scales and encodings; some MI-density and BDM directional
patterns even disagree. They cannot be treated as interchangeable
estimators or quantitative corroboration of the normalised experiment.
Layer-level trends describe displayed structure, not homeostasis or
equilibration. Numerical harmonisation requires the arrays and full
configuration; defining its baseline is not a completed reanalysis.

The network experiment transfers the specified intervention logic and
normalised channels beyond ECA. Variation in those channels and raw mask
descriptions alongside short recurrence illustrates non-equivalence, not
statistical independence or a universal edge-mixture law. Learning,
biological viability and multi-agent objectives are not tested.

Boundary choice is another part of the experiment rather than a
nuisance removed by normalisation. Padding outside the preserved causal
ancestry and matched role exchange test well-defined invariances.
Changing the width of a region governed by a different rule also changes
the physical update map. The resulting directionality reversals
therefore demonstrate architecture dependence; they do not isolate the
effect of an observer repartitioning fixed dynamics. The
mutual-perturbation perspective \cite{abrahao2022observers} motivates
testing the latter question separately.

Appendix~\ref{appendix-dynamical-programme} develops the proposed
organisational-persistence and interaction-timescale tests.
A possible future summary of boundary robustness uses a justified family
\(\mathcal B\) of partitions of the same fixed mechanism and a matched
context ensemble. Let \(\Phi_b\) contain prespecified primitive and
within-context descriptors and let \(\bar\Phi\) be a stated reference
profile. One illustrative summary is
\[
R_{\mathcal B}=1-\frac{1}{|\mathcal B|}
\sum_{b\in\mathcal B}d(\Phi_b,\bar\Phi),
\qquad 0\le d\le1.
\]
The normalisation, weights and treatment of undefined ratios must be
fixed in advance. Including both primitive scores and algebraically
dependent composites can double-weight the same evidence. High
robustness is not sufficient for individuality: an inert system may be
uniformly unresponsive under every partition. Boundary robustness must
therefore be combined with nontrivial organisation and independent
function, and evaluated on held-out contexts if boundaries were selected
using the data. This proposed statistic was not measured here.

Biological applications require an independent viability observable, such
as survival in a specified set over a fixed challenge horizon. Compare
intact and channel-disrupting conditions:
\[
\Delta V=\mathbb E[V\mid\text{intact channel}]
-\mathbb E[V\mid\text{channel disruption}].
\qquad\text{(23)}
\]
A causal interpretation requires a controlled intervention, relevant
marginal-input matching and checks that disruption does not merely damage
the organism or remove essential material resources. A positive
\(\Delta V\) would support a viability-relevant role for the tested
channel, not automatically establish life. This is a proposed bridge to
semantic-information research \cite{kolchinsky2018}, not a result of
the ECA simulations.

Neither dormancy nor passive protection is excluded by the theorem.
A predictable challenge, pre-existing information or one universally
successful response removes its chance-performance premise.
Error-correcting systems can also maintain organisation while suppressing
every response in a narrow perturbation ensemble. Measured absence
therefore concerns the chosen probes and horizon, not an unconditional
absence of agency. Conversely, outward dominance is not required: the
balanced corrective controller regulates successfully at \(D_*=0\).
Raw channels, context structure, persistence and task benefit should be
reported separately before an aggregate receives a functional label.

\subsection{Relation to AID and to information-theoretic
causality}\label{relation-to-aid-and-to-information-theoretic-causality}

AID supplies intervention, counterfactual comparison and response
localisation. Known ECA mechanisms identify effects by construction,
without discovering unknown programs or computing algorithmic mutual
information.

Regular and irregular responses can reach similar recipient fractions
yet have different descriptions. CTM/BDM and DEFLATE provide structural
proxies, not exact Kolmogorov complexity. Analyses of full response profiles
or model changes must declare encodings and approximation limits.
The zero-mask reference calibrates a specific structural contrast; it
does not supply algorithmic conditional information.

Randomised intervention inputs connect the experiment to causal
information theory \cite{lizier2010,ay2008}. The method is not transfer
entropy under a different name, nor does its causal interpretation
require Shannon information to be intrinsically noncausal. The
intervention distribution identifies the experimental channel; mutual
information quantifies that channel. Algorithmic and Shannon observables
answer different questions within a common perturbational methodology.

Ruffini's work raises related questions about persistence, regulation
and algorithmic environmental models
\cite{ruffini2026patterns,ruffini2026regulator}. Its Game of Life
construction is close in experimental spirit to cellular-automaton
studies here. The historical context also includes earlier algorithmic
approaches to life and evolution, Game of Life programmability,
persistent and colliding structures, AID perturbations, open-ended
evolution and observer dependence
\cite{zenil2012,zenilmarshall2013,zenil2015gol,adams2017,hernandez2018,zenil2019gol,abrahao2022observers,zeniladams2022}.
The present contribution is an operational synthesis with explicit
channel definitions, a conditional theorem, counterexamples and a
separation of demonstrated measurements from proposed extensions.

\subsection{Limits of inference and
generalisation}\label{limits-of-inference-and-generalisation}

The simulations use imposed boundaries, a finite rule family, limited
contexts and specific perturbations. The 2,000-pair scan does not average
over all \(2^{210}\) primary initial states. Its bootstrap resamples
complete initial-condition units; it does not cover all choices of
boundary, estimator, encoding or intervention distribution. No living
system is classified, no sensitivity or specificity for life is
estimated, and no universal threshold is proposed. The regulatory
benchmark is not a model of metabolism, reproduction or biological
individuality.

Marginal channels can miss joint information. With independent fair
\(N\), let \(R_1=N\) and \(R_2=N\oplus U\). Each recipient alone has
zero information about \(U\), whereas the pair has one bit. A sum of
marginal information scores cannot recover that synergy. Likewise,
single-bit interventions can miss coordinated state transitions.
Joint interventions and task-aligned recipient representations are
needed when those possibilities matter. Composition identities under
fixed source, recipient and clamping conventions do not turn nonlinear
score cross-terms into evidence of emergence or collective function.

Exact recurrence, stationarity, equilibrium and homeostasis are distinct.
A finite deterministic ECA eventually repeats, but the simulator has no
calibrated temperature, work, heat or entropy-production variables.
Thermodynamic information-flow results require a physical model
\cite{horowitz2014}. Stable BDM, compression or channel curves cannot
supply that missing model. Homeostasis requires a regulated observable
and a restoring mechanism, neither of which follows from a flat curve.

The remaining empirical limitations are explicit. Within-context
reciprocity and episode-level feedback have not been quantified;
zero-referenced numerical BDM results have not been reconstructed;
and the retained exploratory ECA descriptors have not been harmonised
with the primary protocol. These are substantive analyses requiring
their underlying data and configurations, not repairs that can be
achieved by notation alone. They constrain what the figures establish
without invalidating the specified Shannon measurements or the
task-conditional theorem.

\section{Conclusions}\label{conclusions}

This paper develops a perturbational framework connecting causal
information to adaptive regulation without identifying either causal
coupling or complexity with life. In deterministic binary channels,
context-conditioned mutual information has an exact paired-response
representation. Averaging these local quantities produces outward,
inward and clamped internal information-exposure densities. Their
retention-weighted composites satisfy
\[
C^2=D^2+B^2,
\]
an algebraic identity rather than empirical evidence for a life-agency
hierarchy. The reciprocal composite \(B\) concerns ensemble means;
reciprocity within contexts, coordinated episodes and functioning feedback
require separate evidence.

The 2,000-pair scan, detailed ECA cases, recurrence and coding analyses,
geometry controls and signed-network extension demonstrate protocol-dependent
differences among responsiveness, directional imbalance, structural
regularity and recurrence. CTM/BDM add descriptions of encoded objects,
not a second estimate of the same Shannon channel. For paired masks,
the relevant structural contrast subtracts the BDM of a same-shaped
zero mask under identical settings. Existing numerical BDM values remain
explicitly raw, and the exploratory plots remain on their native scales;
neither is presented as a completed baseline-corrected causal analysis.
A common reference preserves matched-estimator comparisons, but raw
positivity alone does not establish an effect.

The strongest theoretical conclusion is conditional and functional.
Successful regulation of the specified unpredictable binary challenge
requires information across each causal cut, while correct and inverted
policies can have identical information profiles and opposite outcomes.
Neither ensemble reciprocity nor a response-mask complexity descriptor
resolves that ambiguity. Information is a necessary resource under the
stated task and causal-chain assumptions, not a sufficient criterion
for agency or a universal measured requirement for every living system.

The proposed extension to life (Appendix~\ref{appendix-dynamical-programme})
therefore combines independent tests:
persistence of a specified organisation, context- or episode-resolved
interaction, causal contribution to viability, and robustness across
justified boundaries of fixed dynamics. Perturbation retention is not
organisational survival, architecture variation is not solely
observational repartitioning, and a stable complexity curve is not
equilibrium. Selective sensing and action may be sequential and need
not produce permanently positive outward dominance.

These distinctions preserve the operational role of information exchange
and algorithmic response analysis while making their limits testable.
They connect the study to earlier work on open-ended evolution and
mutual perturbation \cite{adams2017,abrahao2022observers}, without claiming
that the present automata establish biological function. The framework
delivers measurable causal profiles, a task-specific necessity result
and explicit failures of sufficiency; its broader account of life and
agency is a programme for functional validation, not a scalar verdict.

\section*{Data and code availability}\label{data-and-code-availability}

A reproducibility package containing the complete
simulation and analysis code, explicit configuration and seeds, initial
states for the primary scan and detailed cases, context-level numerical
arrays, plain-text summaries, validation outputs and all
figure-generation scripts are posted. The supplementary package is the data source
for every newly generated numerical claim in this article.

\newpage

\clearpage
\appendix
\numberwithin{figure}{section}
\numberwithin{table}{section}

\section{Collections of subsystems: composition under a fixed experiment}\label{appendix-composition}

Consider a focal subsystem \(A=a\cup b\), with disjoint components \(a,b\) and external environment \(E\). A valid composition law must keep the update mechanism, context distribution, source weighting and clamping intervention fixed. Measuring \(a\) inside \(a\cup b\) is different from declaring \(a\) the entire focal system and clamping \(b\) as part of its environment.

To make the accounting explicit, first use unnormalised exposure totals, with sums rather than averages over source cells. Define
\[
\widetilde X_u=\mathbb E_c\!\left[\sum_{i\in P_u}\sum_{(j,\tau)\in\Omega_{i\to E}(T)}D_{i,j,\tau}(c)\right],\qquad u\in\{a,b\},
\]
and analogously
\[
\widetilde Y_u=\mathbb E_c\!\left[\sum_{i\in P_E}\sum_{(j,\tau)\in\Omega_{i\to u}(T)}D_{i,j,\tau}(c)\right].
\]
For disjoint source and recipient sets,
\[
\widetilde X_{a\cup b}=\widetilde X_a+\widetilde X_b,\qquad
\widetilde Y_{a\cup b}=\widetilde Y_a+\widetilde Y_b.
\]
These are accounting identities for separately indexed intervention--recipient observations, not additivity laws for independent information bits.

For retention, clamp only the common external region \(E\), allowing both \(a\) and \(b\) to evolve. Define
\[
\widetilde S_{u\to v}=\mathbb E_c\!\left[\sum_{i\in P_u}\sum_{(j,\tau)\in\Omega^{\mathrm{late}}_{i\to v}(T)}D^{\mathrm{clamp}\,E}_{i,j,\tau}(c)\right],
\]
where \(u,v\in\{a,b\}\). Then
\[
\widetilde S_{a\cup b}=\widetilde S_{a\to a}+\widetilde S_{b\to b}+\widetilde S_{a\to b}+\widetilde S_{b\to a}.
\]
These terms classify intervention origins and recipient locations. A
within-component term may contain a path that passes through the other
component before returning. It is not an isolated self-retention measurement,
and clamping the other component would change the experiment.
Write \(s_a=\widetilde S_{a\to a}\), \(s_b=\widetilde S_{b\to b}\), \(s_\times=\widetilde S_{a\to b}+\widetilde S_{b\to a}\), and \(L_u^\pm=\widetilde X_u\pm\widetilde Y_u\). With \(\widetilde C_u=s_uL_u^+\) and \(\widetilde D_u=s_uL_u^-\),
\[
\widetilde C_{a\cup b}-(\widetilde C_a+\widetilde C_b)=(L_a^++L_b^+)s_\times+s_bL_a^++s_aL_b^+,
\]
\[
\widetilde D_{a\cup b}-(\widetilde D_a+\widetilde D_b)=(L_a^-+L_b^-)s_\times+s_bL_a^-+s_aL_b^-.
\]
These division-free identities remain valid when a within-component retention term is zero. The last two terms can remain nonzero when \(s_\times=0\); their presence is therefore not evidence of emergent inter-component causation.

Source averaging changes these relationships. If \(x_a,x_b\) are recipient counts averaged over uniformly sampled sources and \(n_u=|P_u|\), then
\[
x_{a\cup b}=\frac{n_ax_a+n_bx_b}{n_a+n_b},
\]
not \(x_a+x_b\). With opportunity normalisation, inward densities must also be combined using eligible-recipient counts. Consequently, the normalised \(X,Y,S\) do not inherit simple total-count additivity. Claims of collective function require an independent functional criterion and an intact-versus-disrupted comparison; joint informational synergy requires a joint intervention distribution and joint-channel analysis.

\clearpage

\section{Extended computational methods}\label{appendix-methods}

\subsection{Hybrid cellular automata and
sampling}\label{hybrid-cellular-automata-and-sampling}

We use a periodic one-dimensional lattice with 210 binary cells. The
central ten cells form \(A\); the remaining 200 form \(E\). Each cell
updates synchronously using its own region's ECA rule and the actual
states of its two neighbours, including neighbours across the boundary.
For rule number \(r_j\), the update is

the rule-table expression in equation (22) of the main text.

which extracts the rule-table bit indexed by the three-cell
neighbourhood. The implementation uses integer right shift followed by a
one-bit mask. This convention is tested against constant, identity,
shift and XOR rules.

The primary scan samples 2,000 distinct ordered pairs \((r_A,r_E)\)
uniformly without replacement from the \(256^2\) possibilities. Sixteen
independent Bernoulli-one-half initial configurations are generated per
pair and evolved for 64 updates before intervention. Every focal cell is
probed. Environment probes occupy the five cells immediately outside
each boundary. The two directions therefore have equal numbers of
one-bit interventions and matched distances from their nearest
interface. The 32-step primary horizon uses lags 17--32 for internal
retention.

The six detailed combinations are \((30,18)\), \((250,30)\), \((30,2)\),
\((30,45)\), \((110,60)\) and \((45,2)\). They retain propagation,
contracting-region, structured-environment and transient cases within a
common protocol. Each is evaluated on 128 independent initial
configurations at ages 0, 16, 64, 128, 256 and 512. The same initial
configurations are followed across ages, enabling paired temporal
comparisons. The cases were specified before the simulations reported
here, but neither the study nor its analyses were preregistered.

\begin{longtable}[]{@{}
  >{\raggedright\arraybackslash}p{(\columnwidth - 2\tabcolsep) * \real{0.2100}}
  >{\raggedright\arraybackslash}p{(\columnwidth - 2\tabcolsep) * \real{0.7900}}@{}}
\caption{Computational design. All simulation settings and initial-state
arrays are included in the reproducibility package.}\tabularnewline
\toprule\noalign{}
\begin{minipage}[b]{\linewidth}\raggedright
Element
\end{minipage} & \begin{minipage}[b]{\linewidth}\raggedright
Specification
\end{minipage} \\
\midrule\noalign{}
\endfirsthead
\toprule\noalign{}
\begin{minipage}[b]{\linewidth}\raggedright
Element
\end{minipage} & \begin{minipage}[b]{\linewidth}\raggedright
Specification
\end{minipage} \\
\midrule\noalign{}
\endhead
\bottomrule\noalign{}
\endlastfoot
Primary ensemble & 2,000 distinct ordered ECA rule pairs; 16 initial
configurations per pair \\
Primary geometry & 10 focal cells and 200 environment cells; periodic
boundary \\
Interventions & One-bit flips; 10 focal and 10 boundary-matched
environment sources \\
Primary timing & Age 64; horizon 32; internal retention evaluated at
lags 17--32 \\
Detailed cases & Six detailed combinations plus five homogeneous
controls; 128 configurations; six ages \\
Geometry sensitivity & Seven geometries; horizons 16, 32, 64; 48
configurations per pair and geometry \\
Uncertainty & 5,000 initial-condition bootstrap resamples; paired
resampling across ages \\
Regulatory benchmark & Eight error settings; 100,000 challenges per
policy and setting \\
\end{longtable}

\subsection{Negative controls and geometry
checks}\label{negative-controls-and-geometry-checks}

Five homogeneous comparisons are added: constant rule 0, identity rule
204, additive rule 90, shift rule 170 and rule 30. These distinguish
extinction, static retention, regular transport and sensitive nonlinear
dynamics. They do not supply biological positive or negative labels.

A sham intervention leaves the source unchanged. A boundary-cut control
replaces each cross-boundary neighbour input by its reference-trajectory
value at every update. The latter preserves the reference trajectory but
prevents a perturbation from being transmitted across the cut. Both
controls are evaluated for all six detailed pairs on 32 contexts each.

The zero-effect assertions for these controls concern paired differences
and Shannon channel scores. For raw BDM, a sham response instead returns the
same-shaped zero-mask reference \(K^0\); its defined structural contrast
\(\Delta K\) is zero. No numerical BDM control results are inferred from the
Shannon checks.

Geometry sensitivity uses \((|A|,|E|)=(6,200)\), \((10,100)\),
\((10,200)\), \((10,400)\), \((20,200)\), \((40,200)\) and \((100,100)\)
and horizons 16, 32 and 64. There are 48 independent initial
configurations per pair and geometry, shared across horizons within that
geometry. At most five sources are sampled at each side of either
interface; when \(|A|=6\), three are used. All available sources are
enumerated, not randomly subsampled. Initial configurations differ
across geometries, so these comparisons concern ensemble means rather
than paired individual trajectories.

Additional exact tests use 50 random contexts for equal-region role
exchange, and 50 for padding a 210-cell configuration to 410 cells while
preserving its relevant initial-state ancestry. The latter are performed
at intervention age zero with horizon 32, not after independently
evolving the two differently sized lattices. A separate scalar
implementation validates the numerical estimator on 24 small-lattice
contexts.

\subsection{Structural descriptors, BDM conventions and recurrence}
\label{structural-descriptors-and-recurrence}

Causal sensitivity, compressibility and recurrence are measured separately.
For each detailed case and age, binary site-frequency entropy and raw
DEFLATE length are calculated from a fixed 64-update unperturbed spacetime
window, treating focal and environmental regions separately. Matrices are
flattened in time-major order, bit-packed most-significant-bit first and
compressed with level-nine raw DEFLATE (negative window-bits setting,
without a zlib header). Code length is reported as compressed bits per
original binary cell. Site-frequency entropy describes the fraction of zeros
and ones, not the entropy rate of the full spatiotemporal process.

DEFLATE is a fixed-code structural descriptor, not an estimate of mutual
information or exact Kolmogorov complexity. Finite-format costs can give
ratios above one. Within-focal-region comparisons across ages use the same
\(64\times10\) representation; focal and environmental code lengths are
not treated as directly interchangeable.

The network analysis describes rectangular XOR masks; exploratory ECA
panels describe layers and native-scale composites. The latter are not
assumed to share the primary scan's clamping, weighting or normalisation.
They support qualitative comparisons, not reconstructed normalised scores
or cross-estimator statistics.

Exact BDM reproduction requires array shape and ordering; software and CTM
table; block dimensions, tiling and remainder handling; and source/context
weights \cite{bdm2018}. Defaults cannot be recovered from plot axes.
Calculate each mask's zero reference under identical settings before
aggregation (Section~\ref{bdm-baseline}). The network summaries supply raw
values, not the numerical reference or corrected results; none are
invented here. A common reference translates means and percentile
intervals without changing between-group differences. Varying shapes or
settings require sample-specific references. This subtraction does not
remove finite-block approximation error. For an unperturbed layer,
a zero-array reference concerns regularity, not causal absence.

We also search for an exact repeat of the complete 210-cell state through
update 4,096. A repeated complete state proves entry into a cycle because
the finite update map is deterministic. The first repeated state's
transient length and cycle period are reported. Failure to observe recurrence
within the window is censored observation, not evidence of aperiodicity,
chaos or thermodynamic nonequilibrium.

\subsection{Regulatory benchmark and statistical
analysis}\label{regulatory-benchmark-and-statistical-analysis}

The binary regulatory benchmark uses (19) at
\((e_s,e_m,e_a)=(0.10,0.05,0.10)\) and at seven equal-error settings
from zero to one-half. For each setting, 100,000 independent challenges
are evaluated for each policy. Corrective and inverting policies share
the challenge and noise realisations, so their paired successes are
complementary. A marginal-matched open-loop control draws the action
independently as a fair bit; it preserves the actuator's input
distribution but destroys the sensor-to-action dependence. Its success
probability is one-half. This is an abstract regulatory task, not a
biological experiment.

For each ECA pair, we first average the context scores \((x_c,y_c,s_c)\)
to obtain \((X,Y,S)\), then calculate \((C,D,B,\alpha)\). We do not
interchange these nonlinear operations. In particular, reported \(B\)
is ensemble reciprocity, not the mean within-context reciprocal score
\(B_{\mathrm{within}}\) or the prevalence \(\pi_{\mathrm{rec}}\). The latter
require the joint context-level scores and are not estimated here. Percentile bootstrap intervals
use 5,000 resamples of complete initial-condition units; cells, time
points and source interventions are not treated as independent
replicates. Temporal intervals resample the same initial conditions
jointly at both ages. Ordinary 95\% intervals are descriptive. For the
three designated transient comparisons \((30,45),(110,60),(45,2)\), the
six endpoint contrasts involving \(C\) and \(D\) additionally use
99.167\% marginal bootstrap intervals, providing a Bonferroni-adjusted
nominal familywise level of 95\%, subject to bootstrap approximation.

Scan sign counts describe estimated pair means, with zero defined by
absolute value at most \(10^{-12}\). They are not exact counts of
population signs after integrating every possible initial state.
Split-half rank correlations compare the first and last eight contexts
per pair and quantify repeatability, not biological validity. For the
benchmark, binomial Wilson intervals accompany individual policy-success
estimates; the policies are not treated as independent when comparing
them.

All numerical experiments use the explicit seeds and software versions
recorded in the accompanying code and configuration files. The
implementation evaluates conditional information exactly within
deterministic contexts; its uncertainty statements concern the sampled
ensembles. No published or manuscript figure was digitised to obtain a
numerical result.

\subsection{Signed-network generalisation using the same perturbational observables}
\label{network-methods}

As a separate proof of concept, the intervention protocol was applied to a
deterministic network of 32 binary nodes partitioned into contiguous
16-node regions on an undirected degree-four ring, with connections to the
two closest nodes on either side. Each undirected edge was independently
assigned symmetric weight \(J_{ij}=J_{ji}\in\{-1,+1\}\), with positive-weight
probability \(p\in\{0,0.25,0.5,0.75,1\}\). For each \(p\), 64 independent
network-and-initial-state contexts were generated. Synchronous dynamics used
\[
h_i(t)=\sum_jJ_{ij}[2z_j(t)-1],
\]
with \(z_i(t+1)=1\) for \(h_i>0\), \(z_i(t+1)=0\) for \(h_i<0\), and
\(z_i(t+1)=z_i(t)\) for ties. Networks evolved for 16 updates before
intervention. The four boundary-connected sources in each region were
probed by single-bit flips. Exposure was normalised over recipient--lag
pairs reachable within graph distance at a 16-update horizon. Retention
used lags 9--16 with the environment clamped to its reference trajectory.

Unperturbed structural measurements comprised binary entropy and level-nine
raw DEFLATE bits per cell over a 32-update window. Complete-network recurrence
was sought over the next 512 updates. The response-mask descriptor was
two-dimensional \texttt{pybdm} BDM of perturbed XOR reference recipient
arrays. Both directions used a full 16-lag by 16-node recipient array,
including zero entries outside actual response support. These rectangular
descriptors differ from the geometrically normalised channel densities.

For the numerical tables and existing figure labels, \(K_X,K_Y\) abbreviate
source- and context-averaged \(K_X^{\mathrm{raw}},K_Y^{\mathrm{raw}}\).
They are not baseline-subtracted quantities. For fixed estimator settings,
let
\[
K_0=\mathrm{BDM}_{\theta}(0_{16\times16}).
\]
The corresponding baseline-aware sum would be
\(\Delta K_X+\Delta K_Y=K_X+K_Y-2K_0\). Its value is not claimed here because
the numerical zero-mask reference is not reported. The same reference
applies across the five mixtures, so comparisons between their raw sums
are unaffected by this constant subtraction. Its absolute size still
cannot be interpreted as independent transmitted information or functional
coupling.

For each \(p\), \(x_c,y_c,s_c\) were averaged over the 64 contexts before
constructing \(C,D,B,\alpha\), matching the ECA convention. This estimates
ensemble products, not \(B_{\mathrm{within}}\), \(\pi_{\mathrm{rec}}\) or the
average of context-level products. Percentile intervals used 5,000
complete-context bootstrap resamples, treating a sampled network and
initial state as one unit. The sign-mixture ensembles were independently
generated, not paired across \(p\). Bootstrap uncertainty concerns context
sampling; it does not quantify uncertainty from BDM block conventions.

\clearpage

\subsection{Paired intervention visualisation}

\begin{figure}[ht]
\centering
\begin{subfigure}{0.32\textwidth}
\includegraphics[width=\linewidth]{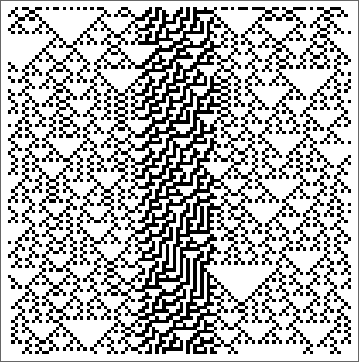}
\caption{Unperturbed control case.}
\end{subfigure}
\begin{subfigure}{0.32\textwidth}
\includegraphics[width=\linewidth]{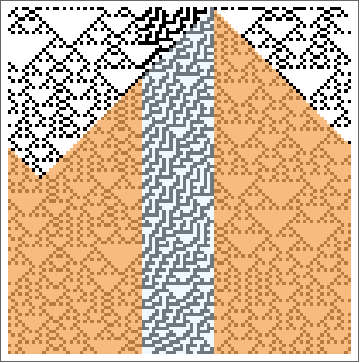}
\caption{Perturbation of the focal subsystem.}
\end{subfigure}
\begin{subfigure}{0.32\textwidth}
\includegraphics[width=\linewidth]{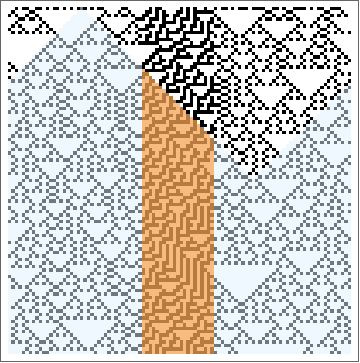}
\caption{Perturbation of the environment.}
\end{subfigure}
\caption{Perturbational visualisation for rule 30 in a rule-18
environment. The panels show the control trajectory and perturbations
originating on either side of the imposed boundary. They are retained to
make explicit the common intervention logic underlying both the Shannon
and structural analyses. The panels alone do not specify a BDM estimator,
zero-mask reference or functional criterion.}
\end{figure}

\clearpage

\section{Extended ECA results and controls}\label{appendix-extended-results}

\subsection{Detailed cases distinguish outward effects from
retained
coupling}\label{detailed-cases-distinguish-outward-effects-from-retained-coupling}

\begin{longtable}[]{@{}
  >{\raggedright\arraybackslash}p{(\columnwidth - 12\tabcolsep) * \real{0.1300}}
  >{\raggedright\arraybackslash}p{(\columnwidth - 12\tabcolsep) * \real{0.0800}}
  >{\raggedright\arraybackslash}p{(\columnwidth - 12\tabcolsep) * \real{0.0800}}
  >{\raggedright\arraybackslash}p{(\columnwidth - 12\tabcolsep) * \real{0.0800}}
  >{\raggedright\arraybackslash}p{(\columnwidth - 12\tabcolsep) * \real{0.3500}}
  >{\raggedright\arraybackslash}p{(\columnwidth - 12\tabcolsep) * \real{0.1400}}
  >{\raggedright\arraybackslash}p{(\columnwidth - 12\tabcolsep) * \real{0.1400}}@{}}
\caption{Primitive and derived scores at intervention age 64. Means are
over 128 independent initial configurations; intervals resample those
configurations. A ratio of one with zero coupling, as for 250 / 30, is
not evidence of retained agency.}\tabularnewline
\toprule\noalign{}
\begin{minipage}[b]{\linewidth}\raggedright
Rules A / E
\end{minipage} & \begin{minipage}[b]{\linewidth}\raggedright
\(X\)
\end{minipage} & \begin{minipage}[b]{\linewidth}\raggedright
\(Y\)
\end{minipage} & \begin{minipage}[b]{\linewidth}\raggedright
\(S\)
\end{minipage} & \begin{minipage}[b]{\linewidth}\raggedright
\(C\) {[}95\% interval{]}
\end{minipage} & \begin{minipage}[b]{\linewidth}\raggedright
\(D\)
\end{minipage} & \begin{minipage}[b]{\linewidth}\raggedright
\(\alpha\)
\end{minipage} \\
\midrule\noalign{}
\endfirsthead
\toprule\noalign{}
\begin{minipage}[b]{\linewidth}\raggedright
Rules A / E
\end{minipage} & \begin{minipage}[b]{\linewidth}\raggedright
\(X\)
\end{minipage} & \begin{minipage}[b]{\linewidth}\raggedright
\(Y\)
\end{minipage} & \begin{minipage}[b]{\linewidth}\raggedright
\(S\)
\end{minipage} & \begin{minipage}[b]{\linewidth}\raggedright
\(C\) {[}95\% interval{]}
\end{minipage} & \begin{minipage}[b]{\linewidth}\raggedright
\(D\)
\end{minipage} & \begin{minipage}[b]{\linewidth}\raggedright
\(\alpha\)
\end{minipage} \\
\midrule\noalign{}
\endhead
\bottomrule\noalign{}
\endlastfoot
30 / 18 & 0.260 & 0.362 & 0.416 & 0.259 {[}0.251, 0.268{]} & -0.042 &
-0.164 \\
250 / 30 & 0.059 & 0 & 0 & 0 {[}0, 0{]} & 0 & 1.000 \\
30 / 2 & 0.134 & 0.101 & 0.299 & 0.070 {[}0.066, 0.074{]} & 0.010 &
0.139 \\
30 / 45 & 0.311 & 0.353 & 0.388 & 0.258 {[}0.248, 0.268{]} & -0.016 &
-0.063 \\
110 / 60 & 0.026 & 0.148 & 0.326 & 0.057 {[}0.054, 0.059{]} & -0.040 &
-0.699 \\
45 / 2 & 0.024 & 0.035 & 0.272 & 0.016 {[}0.015, 0.017{]} & -0.003 &
-0.197 \\
\end{longtable}

The rule-250 focal region has nonzero outward exposure, \(X=0.059\), but
vanishing late internal retention and inward exposure at age 64.
Consequently its coupling is zero even though its unweighted
directionality ratio is one. This case demonstrates why a large
directional ratio alone is not evidence of internally sustained
organisation.

\begin{figure}[ht]
\centering
\includegraphics[width=2in]{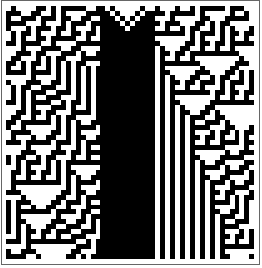}
\caption{Cellular-automaton evolution for rule 250 as the focal subsystem in a rule-30 environment. }
\end{figure}

Rule 30 in rule 2 has positive aggregate reciprocity \(B\) and modest
outward dominance: its mean inward, outward and retention scores are
positive. This does not establish reciprocal transmission within every
context, or report how many contexts have all three positive scores. Rule 30 in rule 45
shows larger coupling but slight inward dominance. Rule 110 in rule 60
is more strongly inward dominated. These are descriptions of causal
responsiveness under specific boundary probes, not rankings of which
automaton is more alive.

\begin{figure}[ht]
\centering
\includegraphics[width=2in]{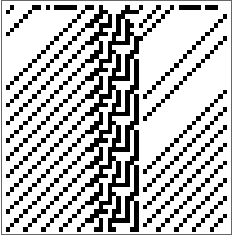}
\caption{Cellular-automaton evolution for rule 30 as the focal subsystem in a rule-2 environment.}
\end{figure}

The homogeneous controls further delimit interpretation. Rule 0 has
\(X=Y=S=C=0\). Rule 204 retains the altered focal bit, giving \(S=0.1\),
but transmits nothing across the boundary, so \(C=0\). Rule 170
transports perturbations across the boundary but retains none within the
focal region over the late window, again giving \(C=0\). The additive
rule-90 lattice has \(C=0.1226\) despite having no implemented
regulatory objective. Homogeneous rule 30 has \(C=0.2784\) and
\(\alpha=-0.0866\): even identical update rules on both sides do not
force zero asymmetry when the regions have unequal geometry.

\subsection{Transient changes in coupling do not imply monotonic
changes in
directionality}\label{transient-changes-in-coupling-do-not-imply-monotonic-changes-in-directionality}

\begin{longtable}[]{@{}
  >{\raggedright\arraybackslash}p{(\columnwidth - 4\tabcolsep) * \real{0.1600}}
  >{\raggedright\arraybackslash}p{(\columnwidth - 4\tabcolsep) * \real{0.4200}}
  >{\raggedright\arraybackslash}p{(\columnwidth - 4\tabcolsep) * \real{0.4200}}@{}}
\caption{Paired changes from intervention age zero to 512. These are
differences between aggregate scores constructed from primitive ensemble
means, not averages of per-context aggregate-score differences.
Multiplicity-adjusted intervals for the three designated transient
comparisons are stored in the numerical summary.}\tabularnewline
\toprule\noalign{}
\begin{minipage}[b]{\linewidth}\raggedright
Rules A / E
\end{minipage} & \begin{minipage}[b]{\linewidth}\raggedright
\(\Delta C\) {[}95\% interval{]}
\end{minipage} & \begin{minipage}[b]{\linewidth}\raggedright
\(\Delta D\) {[}95\% interval{]}
\end{minipage} \\
\midrule\noalign{}
\endfirsthead
\toprule\noalign{}
\begin{minipage}[b]{\linewidth}\raggedright
Rules A / E
\end{minipage} & \begin{minipage}[b]{\linewidth}\raggedright
\(\Delta C\) {[}95\% interval{]}
\end{minipage} & \begin{minipage}[b]{\linewidth}\raggedright
\(\Delta D\) {[}95\% interval{]}
\end{minipage} \\
\midrule\noalign{}
\endhead
\bottomrule\noalign{}
\endlastfoot
30 / 18 & 0.0207 {[}0.0076, 0.0334{]} & -0.0245 {[}-0.0316,
-0.0172{]} \\
250 / 30 & 0 {[}0, 0{]} & 0 {[}0, 0{]} \\
30 / 2 & -0.0030 {[}-0.0082, 0.0022{]} & 0.0070 {[}0.0025, 0.0114{]} \\
30 / 45 & -0.0100 {[}-0.0226, 0.0031{]} & 0.0035 {[}-0.0012,
0.0080{]} \\
110 / 60 & -0.0191 {[}-0.0228, -0.0154{]} & -0.0009 {[}-0.0036,
0.0018{]} \\
45 / 2 & -0.0294 {[}-0.0334, -0.0255{]} & 0.0012 {[}-0.0013,
0.0037{]} \\
\end{longtable}

For rule 110 in rule 60, coupling falls from 0.0749 to 0.0558 between
ages zero and 512. For rule 45 in rule 2, it falls from 0.0442 to
0.0147. Both decreases remain separated from zero by the
multiplicity-adjusted bootstrap intervals. Their signed directional
changes, however, are not separated from zero by ordinary 95\%
intervals. In particular, a negative score moving closer to zero would
be an increase in the signed score and a decrease in its magnitude;
these must not be conflated.

For rule 30 in rule 45, the coupling difference is -0.0100 with an
interval spanning zero, and the signed directional difference also spans
zero. These data support the absence of a clearly resolved endpoint
change at the available precision, not a proof of stationarity or
equilibrium. For rule 30 in rule 18, coupling instead increases. There
is no general monotonic law linking age, complexity and coupling across
the six examples.

\begin{figure}[ht]
\centering
\includegraphics[width=2in]{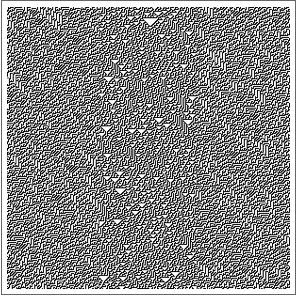}
\caption{Cellular-automaton evolution for rule 30 as the focal subsystem in a rule-45 environment. }
\end{figure}

\begin{figure}[ht]
\centering
\includegraphics[width=2in]{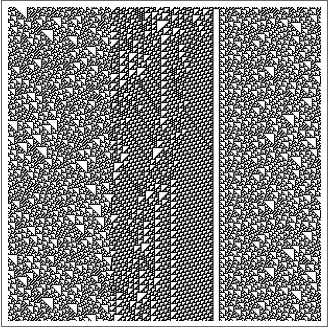}
\caption{Cellular-automaton evolution for rule 110 as the focal subsystem in a rule-60 environment. }
\end{figure}

\begin{figure}[ht]
\centering
\includegraphics[width=2in]{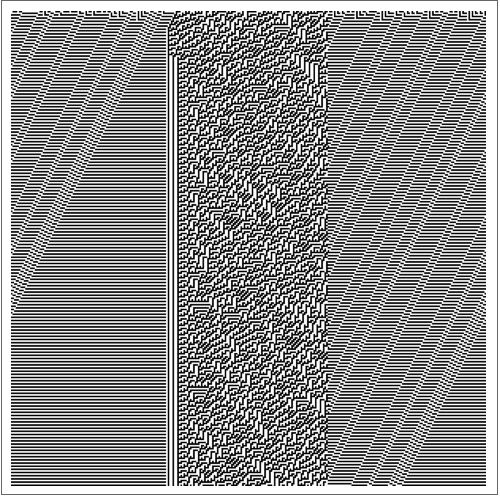}
\caption{Cellular-automaton evolution for rule 45 as the focal subsystem in a rule-2 environment. }
\end{figure}

The aggregate coupling and directionality time courses are displayed in
the main text; the table above retains every paired endpoint contrast.

\subsection{Exact recurrence and structural regularity are
different from causal
responsiveness}\label{exact-recurrence-and-structural-regularity-are-different-from-causal-responsiveness}

\begin{longtable}[]{@{}
  >{\raggedright\arraybackslash}p{(\columnwidth - 8\tabcolsep) * \real{0.1400}}
  >{\raggedright\arraybackslash}p{(\columnwidth - 8\tabcolsep) * \real{0.2000}}
  >{\raggedright\arraybackslash}p{(\columnwidth - 8\tabcolsep) * \real{0.2300}}
  >{\raggedright\arraybackslash}p{(\columnwidth - 8\tabcolsep) * \real{0.2300}}
  >{\raggedright\arraybackslash}p{(\columnwidth - 8\tabcolsep) * \real{0.2000}}@{}}
\caption{Exact recurrence of the complete lattice state through update
4,096. A dash denotes no detected recurrence, not evidence of an
infinite nonperiodic trajectory. Transient and period summaries are
conditional on recurrence being observed.}\tabularnewline
\toprule\noalign{}
\begin{minipage}[b]{\linewidth}\raggedright
Rules A / E
\end{minipage} & \begin{minipage}[b]{\linewidth}\raggedright
Observed cycles
\end{minipage} & \begin{minipage}[b]{\linewidth}\raggedright
Median transient
\end{minipage} & \begin{minipage}[b]{\linewidth}\raggedright
Median period
\end{minipage} & \begin{minipage}[b]{\linewidth}\raggedright
Period range
\end{minipage} \\
\midrule\noalign{}
\endfirsthead
\toprule\noalign{}
\begin{minipage}[b]{\linewidth}\raggedright
Rules A / E
\end{minipage} & \begin{minipage}[b]{\linewidth}\raggedright
Observed cycles
\end{minipage} & \begin{minipage}[b]{\linewidth}\raggedright
Median transient
\end{minipage} & \begin{minipage}[b]{\linewidth}\raggedright
Median period
\end{minipage} & \begin{minipage}[b]{\linewidth}\raggedright
Period range
\end{minipage} \\
\midrule\noalign{}
\endhead
\bottomrule\noalign{}
\endlastfoot
30 / 18 & 0 / 128 & - & - & - \\
250 / 30 & 128 / 128 & 515 & 2 & 1--2 \\
30 / 2 & 128 / 128 & 397 & 4 & 3--8 \\
30 / 45 & 0 / 128 & - & - & - \\
110 / 60 & 128 / 128 & 44.5 & 256 & 256--512 \\
45 / 2 & 128 / 128 & 204 & 3 & 3--6 \\
\end{longtable}

Exact complete-state cycles are detected in every tested initial
condition for four of the six detailed pairs. For rule 110 in rule 60,
the median transient is only 44.5 updates, but the median cycle period
is 256 updates. A short local window can therefore look irregular even
after exact recurrence has been reached. Neither high short-window
entropy nor stable code length establishes absence of periodic
organisation.

Rule 30 in rule 2 provides a clear separation of pattern regularity and
causal response. Its mean agent binary entropy remains close to one bit
per site (0.9981 at age zero and 0.9992 at age 512), while the 64-step
DEFLATE ratio falls from 0.6183 to 0.1396. Coupling changes only from
0.0757 to 0.0727, with its paired confidence interval spanning zero. The
system can therefore become much more compressible while maintaining a
comparable response to intervention. This is not a paradox: the code
measures the regularity of the unperturbed spacetime array, whereas
\(C\) measures sensitivity to controlled changes.

For rule 110 in rule 60, the same short-window code ratio changes much
less, from 0.9169 to 0.9040, even though coupling decreases. The two
cases rule out a simple identification of coding complexity with causal
coupling in either direction. We make no claim that these fixed-code
results reproduce a BDM estimate or measure algorithmic mutual
information.

\subsection{Geometry and intervention
controls}\label{geometry-and-intervention-controls}

Geometry can alter both the magnitude and the sign of directionality.
With a 200-cell environment and a 32-step horizon, rule 45 in rule 2
changes from \(\alpha=-0.438\) at focal width six to -0.238 at width ten
and +0.115 at width twenty. Rule 30 in rule 2 changes from -0.077 at
width six to +0.151 at width ten. Therefore dimensionless normalisation
is not a licence to treat the score as an intrinsic, scale-free property
of a rule pair.

Changing focal width changes the allocation of update rules and therefore
the update map as well as the intervention geometry. These reversals
demonstrate architecture-dependent causal attribution, not a change in
agency caused solely by an observer moving a boundary around an otherwise
unchanged system. The mutual-perturbation framework makes the
observer--system decomposition explicit \cite{abrahao2022observers}, but
a direct test of observational-boundary robustness would keep the physical
update map, underlying trajectories and context distribution fixed while
recomputing matched profiles for several justified partitions. That
separate test has not been performed here.

In contrast, adding distant cells while preserving the relevant local
initial-state ancestry leaves the scores exactly unchanged in all 50
padding tests. Equal-region role exchange swaps \(X\) and \(Y\) within a
maximum discrepancy of \(1.12\times10^{-16}\) in 50 tests. This
distinguishes desirable invariances of a fixed experiment from genuine
changes to the experiment itself.

Every sham intervention yields zero Shannon effect scores. This does not
mean a zero raw BDM: the corresponding unchanged-response mask has
\(\mathrm{BDM}_{\theta}(0)=K^0\), and only its defined matched contrast
is zero. Every boundary-cut
control yields exactly zero \(X\) and \(Y\), across 192 pair-context
combinations. The independent scalar implementation agrees exactly with
the accelerated implementation in all 24 checks. Across the scan, the
largest factorisation residual is \(2.78\times10^{-17}\) and no bound
violations occur.

Clamping also changes interpretation. At age 64,
\(S_{\mathrm{open}}=0.4440\) and \(S=0.3884\) for rule 30 in rule 45,
whereas for rule 45 in rule 2 the corresponding values are 0.1716 and
0.2717. Clamping need not reduce the response: environmental feedback
can either sustain or suppress a difference. Accordingly,
\(S_{\mathrm{open}}-S\) is not assumed to be a nonnegative decomposition
of causal influence.

\begin{figure}[ht]
\centering
\includegraphics[width=4in,height=2.5in]{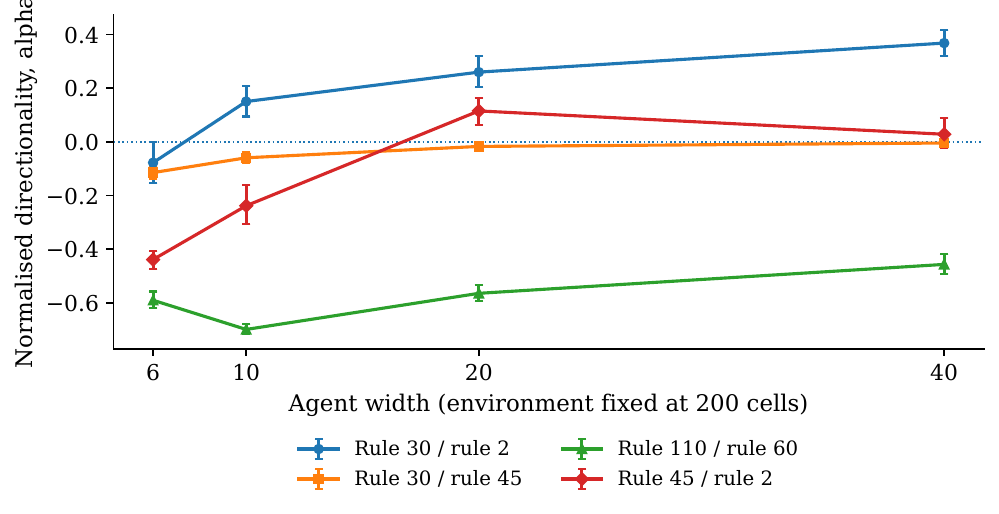}
\caption{Directionality depends on the chosen focal width. The
environment has 200 cells and the horizon is 32 updates. Error bars are
95\% initial-condition bootstrap intervals from 48 configurations per
pair and width. The two regions of a partition cannot be treated as
interchangeable merely because both scores are densities; the primary
effect of this plot is to expose, not eliminate, scale dependence.}
\end{figure}

\clearpage

\section{Complete signed-network estimates}\label{appendix-network-table}

\begin{table}[H]
\centering
\scriptsize
\setlength{\tabcolsep}{4pt}
\caption{Signed-network ensemble summaries with 95\% context-bootstrap
intervals (64 contexts per mixture). \(K_X+K_Y\) abbreviates the sum of
raw BDM descriptors; the zero-mask offset \(2K_0\) has not been subtracted.
These values are not channel mutual information. The table does not
estimate within-context reciprocity.}
\label{tab:multiagent-aggregate}
\begin{tabular}{@{}cccc@{}}
\toprule
$p$ & $C$ [95\% CI] & $D$ [95\% CI] & $K_X+K_Y$ [95\% CI]\\
\midrule
0.00 & 0.001750 [0.001353, 0.002223] & $-0.000062$ [$-0.000363$, 0.000204] & 70.45 [67.93, 72.96]\\
0.25 & 0.001421 [0.000972, 0.001983] & $-0.000117$ [$-0.000396$, 0.000114] & 69.51 [65.51, 73.95]\\
0.50 & 0.001118 [0.000744, 0.001544] & 0.000005 [$-0.000228$, 0.000199] & 67.00 [63.42, 70.81]\\
0.75 & 0.001054 [0.000701, 0.001522] & 0.000015 [$-0.000227$, 0.000211] & 64.06 [61.05, 67.35]\\
1.00 & 0.000749 [0.000508, 0.001045] & $-0.000119$ [$-0.000332$, 0.000059] & 60.28 [58.16, 62.43]\\
\bottomrule
\end{tabular}
\end{table}

\clearpage

\section{Exploratory ECA information and BDM comparisons}
\label{exploratory-results}

The retained panels examine related ECA examples using information-density
and BDM constructions in their native units. They preserve the structural
comparisons that motivated the present separation of causal responsiveness,
descriptional structure and function. They are not additional measurements
of the normalised \(C,D\) in equation (12).

This distinction is visible in the axes: the exploratory MI scan is labelled
nonnormalised and its values exceed \(C\le2\); the BDM composites have still
larger numerical scales. In this subsection the plotted \(C,D\) are denoted
\(C_{\mathrm{exp}},D_{\mathrm{exp}}\). The original image labels are retained,
but they do not identify the plotted values with the primary causal scores.

The panels do not specify enough of the intervention ensemble, BDM
configuration, clamping and aggregation for a matched quantitative
reconstruction. Their visual observations are therefore kept separate
from the context-array results, without inferred correlations or intervals.

\subsection{Native-scale rule-pair comparisons}

Associations between sum-like and difference-like composites of common
nonnegative components can arise algebraically. They do not show that
life causes agency or that both estimators measure the same quantity.
Cross-estimator comparisons require matched contexts and explicit formulas.

Zero-mask offsets must pass through the actual BDM aggregation:
subtracting a constant from a product is not equivalent to replacing its
primitive terms with structural contrasts. Neither composites nor signs
are retrospectively recalibrated from image pixels.

\begin{figure}[ht]
\centering
\includegraphics[width=0.68\textwidth]{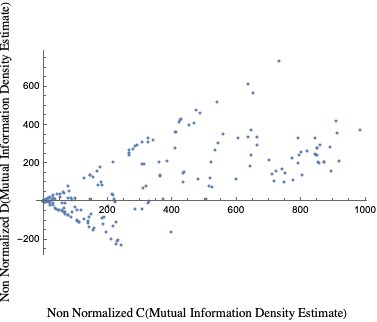}
\caption{Retained exploratory MI-density scan in native nonnormalised units.
The original \(C,D\) labels denote \(C_{\mathrm{exp}},D_{\mathrm{exp}}\), not
the normalised scores of the primary 2,000-pair scan. No numerical conversion
between the constructions is inferred from this panel.}
\label{fig:exploratory-mi-scan}
\end{figure}

\begin{figure}[ht]
\centering
\includegraphics[width=0.68\textwidth]{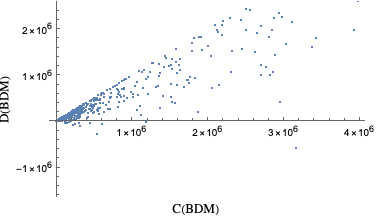}
\caption{Retained exploratory BDM composites in native units. The displayed
values are not baseline-corrected channels or algorithmic mutual information.
Zero-mask references would have to be applied to the actual primitive
objects and aggregation formula before interpreting structural contrasts.}
\label{fig:exploratory-bdm-scan}
\end{figure}

\subsection{Contrasting focal and environmental dynamics}

The cases contrast a simple focal rule in a complex environment with a
complex focal rule in a more structured environment. Wolfram classes here
describe dynamical examples, not biological classes.

For rule 250 in rule 30, the displayed MI distributions for
\(C_{\mathrm{exp}}\) and \(D_{\mathrm{exp}}\) overlap substantially and are
positive; the BDM distributions also overlap. They do not support describing
directionality as uniformly weak relative to coupling in that example.
These distributions are distinct from the age-64 primary result
\(X=0.059,\ Y=S=C=D=0\). In that protocol, the retention multiplier makes
coupling zero despite outward exposure. Positive native-scale exploratory
values cannot be presented as replication of the normalised result.

For rule 30 in rule 2, the exploratory MI panel separates a large
coupling-like distribution from a directional distribution near zero. The
BDM panel displays its own native-scale distribution. These observations
describe the plotted constructions; neither establishes functional
regulation or measured within-context reciprocity.

\begin{figure}[ht]
\centering
\includegraphics[width=0.68\textwidth]{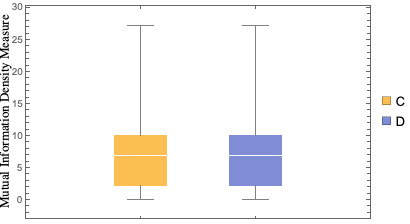}
\caption{Exploratory MI-density distributions for rule 250 in rule 30.
Native-scale \(C_{\mathrm{exp}},D_{\mathrm{exp}}\) overlap. They are not the
normalised age-64 scores, for which zero retention gives \(C=D=0\).}
\end{figure}

\begin{figure}[ht]
\centering
\includegraphics[width=0.68\textwidth]{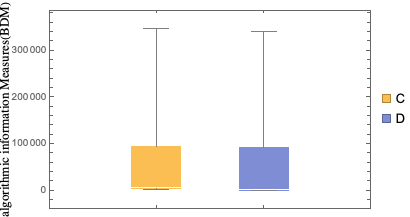}
\caption{Exploratory BDM composites for rule 250 in rule 30, without a
reported zero-mask correction. Their absolute positivity is not evidence
of positive retained causal coupling.}
\end{figure}

\begin{figure}[ht]
\centering
\includegraphics[width=0.68\textwidth]{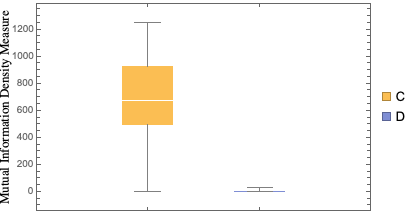}
\caption{Exploratory MI-density distributions for rule 30 in rule 2.
The native-scale separation is distinct from the primitive means and
normalised directionality reported in the primary analysis.}
\end{figure}

\begin{figure}[ht]
\centering
\includegraphics[width=0.68\textwidth]{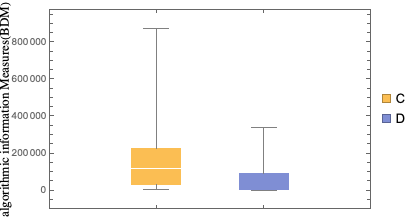}
\caption{Exploratory BDM composites for rule 30 in rule 2. These are raw
structural composites, not zero-baseline causal scores. Comparing estimators
quantitatively would require matched definitions, settings and contexts.}
\end{figure}

\subsection{Structural descriptions through dynamical transients}

The layer-level panels describe unperturbed focal and environmental arrays.
They are distinct from the network XOR masks. Within-layer temporal
comparisons use the plotted representation; absolute cross-layer
comparisons remain sensitive to size and decomposition. An all-zero
reference for a layer describes structural regularity, not absence of
causal transmission.

For rule 30/45, both layer BDM curves fluctuate around comparatively stable
levels. The native MI directional curve is largely negative, whereas the
native BDM directional curve is largely positive. This visible sign
difference rules out treating the panels as interchangeable measurements
of the same signed causal quantity. A stable envelope does not establish
equilibrium.

For rule 110/60, focal-layer BDM decreases while the environmental curve
remains comparatively stable. The BDM coupling-like composite falls, while
its negative directional composite approaches zero: an increase in signed
value and a decrease in magnitude. The separately specified normalised
analysis resolves a decrease in \(C\), but not an endpoint change in signed
\(D\) at the reported precision.

For rule 45/2, focal BDM shows a pronounced stepped decline; environmental
BDM drops early and subsequently remains more stable. The composites also
change, with directional values generally moving towards zero. The two
layers should not be described as showing identical sustained declines.
These are descriptive observations of structural regularisation, not
statistical tests based on digitised trajectories.

\begin{figure}[ht]
\centering
\includegraphics[width=0.68\textwidth]{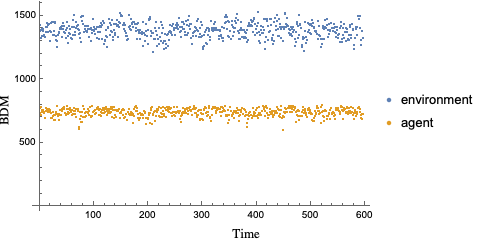}
\caption{Retained layer-level BDM for rule 30 in rule 45. Both layers
fluctuate around comparatively stable levels. The historical asset filename
does not establish thermodynamic equilibrium; these are unperturbed-layer
descriptors rather than response-mask contrasts.}
\label{fig:legacy3045layers}
\end{figure}

\begin{figure}[ht]
\centering
\includegraphics[width=0.68\textwidth]{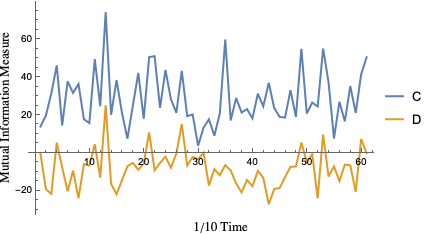}
\caption{Exploratory MI-density time series for rule 30 in rule 45.
The original units and time convention are retained. These are not the
ensemble-normalised time courses from the primary protocol.}
\label{fig:legacy3045mi}
\end{figure}

\begin{figure}[ht]
\centering
\includegraphics[width=0.68\textwidth]{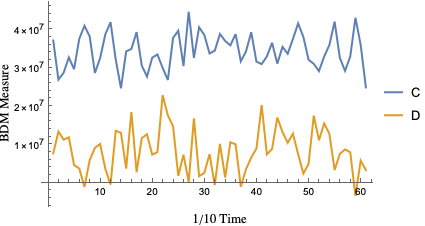}
\caption{Exploratory BDM composites for rule 30 in rule 45. The predominantly
positive directional curve differs from the predominantly negative
MI-density curve in Figure~\ref{fig:legacy3045mi}. No unreported baseline
correction is assumed. The corresponding CA spacetime example is retained
in the preceding results.}
\end{figure}

\begin{figure}[ht]
\centering
\includegraphics[width=0.68\textwidth]{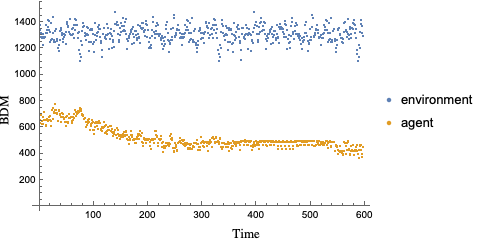}
\caption{Layer-level BDM for rule 110 in rule 60. Focal structural description
length decreases under the plotted representation while environmental BDM
remains comparatively stable. No thermodynamic state is inferred.}
\end{figure}

\begin{figure}[ht]
\centering
\includegraphics[width=0.68\textwidth]{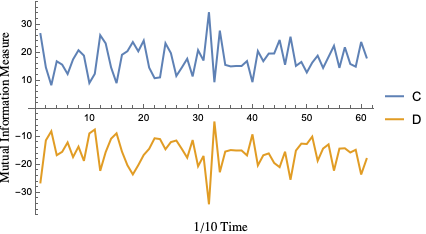}
\caption{Exploratory MI-density composites for rule 110 in rule 60. Native
values and fluctuations do not replace the normalised paired endpoint
estimates or their context-bootstrap intervals.}
\end{figure}

\begin{figure}[ht]
\centering
\includegraphics[width=0.68\textwidth]{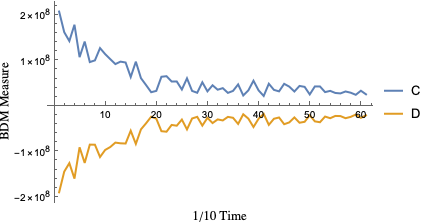}
\caption{Exploratory BDM composites for rule 110 in rule 60. The coupling-like
curve declines and negative directionality approaches zero. Decreasing
magnitude is not decreasing signed value. These native composites are not
reported as baseline-corrected responses.}
\end{figure}

\begin{figure}[ht]
\centering
\includegraphics[width=0.68\textwidth]{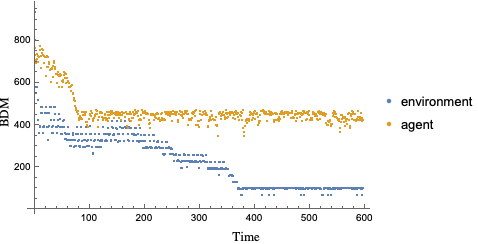}
\caption{Layer-level BDM for rule 45 in rule 2. The focal curve shows stepped
simplification; the environmental curve drops early and then stabilises more
closely. The layers do not exhibit identical temporal changes.}
\end{figure}

\begin{figure}[ht]
\centering
\includegraphics[width=0.68\textwidth]{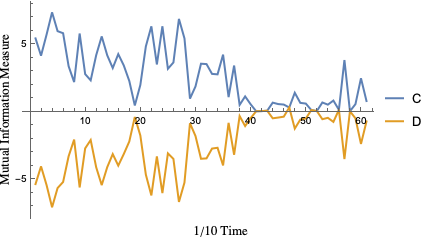}
\caption{Exploratory MI-density composites for rule 45 in rule 2. Their
native-scale changes are descriptive observations under that construction,
distinct from the normalised ensemble comparison.}
\end{figure}

\begin{figure}[h]
\centering
\includegraphics[width=0.68\textwidth]{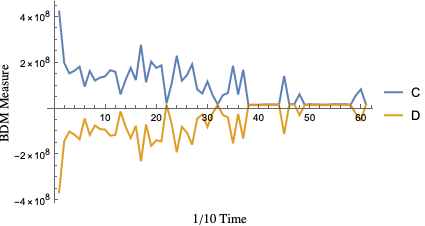}
\caption{Exploratory BDM composites for rule 45 in rule 2. Structural
regularisation and composite changes do not establish functional loss,
baseline-corrected causal transmission or a biological classification.}
\end{figure}

The structural panels preserve useful comparisons but do not establish
statistical independence, a universal complexity trend or equivalence of
estimators. Quantitative claims about the normalised causal profile rest on
its specified simulations; the DEFLATE and recurrence results provide
separate numerical dissociations; the task construction establishes the
functional distinction. A harmonised BDM comparison would require the
stored trajectories, exact estimator settings and matched zero references.

\clearpage

\section{Proposed dynamical classification of life-like causal organisation}\label{appendix-dynamical-programme}

The results motivate an operational research programme, not a
classification of the simulated automata as living or nonliving.
Permanently positive outward dominance is not required by the regulation
theorem. Sensing and acting can involve different directions of influence,
and a successful controller can have zero directional imbalance. Three
questions must instead be kept separate: whether an organisation persists,
whether its boundary supports appropriately coordinated interaction, and
whether that interaction contributes to an independently defined function.

Persistence of a perturbation is not persistence of an organisation.
The measured \(S\) describes a late, clamped internal response to a specified
probe; an error-correcting system can preserve its organisation by erasing
that response. To measure organisational persistence, specify an observable
\(O(A_t)\), an admissible set \(\mathcal V_O\), and a context ensemble,
then estimate a survival function such as
\[
L_A(\tau)=
\Pr\!\left[O(A_{t+u})\in\mathcal V_O
\text{ for every }0\le u\le\tau
\,\middle|\, O(A_t)\in\mathcal V_O\right].
\]
For a prespecified survival level \(\rho\), one possible timescale is
\(\tau_A(\rho)=\sup\{\tau:L_A(\tau)\ge\rho\}\). The observable, threshold,
sampling interval and censoring horizon are part of the definition.
Choosing a trivial admissible set would make persistence uninformative.

A separate causal-persistence experiment could randomly prepare an
internal feature using an intervention input \(U_O\) and measure
\[
P_A^O(\tau)=I(U_O;O(A_{t+\tau})\mid\mathcal C_{\mathrm{ctx}}),
\]
where \(\mathcal C_{\mathrm{ctx}}\) contains only the specified
pre-intervention context and the environmental-control protocol is stated.
This is a channel about the prepared feature, not a survival probability.
Its decay time need not equal \(\tau_A\): stable organisation can actively
remove the preparation. Neither \(L_A\) nor \(P_A^O\) has been measured
in the present experiments.

For interaction, define \(\tau_X\) and \(\tau_Y\) as the first lags
at which outward and inward effects meet prespecified detection criteria.
If the hypothesis concerns access in \emph{both} directions, the relevant
completion time is
\[
\tau_{\partial}=\max\{\tau_X,\tau_Y\},
\qquad
\Lambda=\frac{\tau_A(\rho)}{\tau_{\partial}}.
\]
The minimum would time only the faster direction and could misleadingly
classify a one-way channel as rapid reciprocal access. A direction not
detected within the observation window must be reported as undetected or
censored; it must not be assigned the other direction's latency. Estimates
of \(\Lambda\) require compatible time units, finite positive denominators
and a sufficiently long observation window. No value of this ratio is
inferred from the existing fixed-window \(S\) or recurrence measurements.

A large \(\Lambda\), where estimable, would describe relatively persistent
organisation with faster two-way causal access. It would not by itself
establish selective or beneficial interaction. For that purpose, use
matched contexts and report both the aggregate
\(B=2S\sqrt{XY}\) and a within-context summary such as
\(B_{\mathrm{within}}=\mathbb E[2s_c\sqrt{x_cy_c}]\), together with
\(\pi_{\mathrm{rec}}\). Positive aggregate \(B\) can arise by mixing
one-way contexts. Even positive within-context scores establish two
separately probed directions, not necessarily a simultaneously operating
feedback loop. Where sensing and action occur sequentially, define the
episode, lag ordering and task relevance before testing their coordination.

The following regimes organise these questions; their qualitative
thresholds are not fitted diagnostic criteria.

\begin{center}
\begin{tabular}{
    >{\raggedright\arraybackslash}p{0.18\textwidth}
    >{\raggedright\arraybackslash}p{0.18\textwidth}
    >{\raggedright\arraybackslash}p{0.18\textwidth}
    >{\raggedright\arraybackslash}p{0.31\textwidth}
}
\hline
\textbf{Regime} & \textbf{Persistence of organisation} &
\textbf{Boundary interaction} & \textbf{Interpretation} \\
\hline
Decoupled or inert & Low & Low &
Little measured persistence or cross-boundary response \\
Persistent isolated & High & Low &
Persistent structure; maintenance mechanism remains to be tested \\
Inward-driven & Assessed separately & \(Y\gg X\) &
Inward-dominated response, not necessarily absence of agency \\
Outward-led & Assessed separately & \(X\gg Y\) &
Source or actuator; task benefit is undetermined \\
Highly coupled & Assessed separately & High \(X,Y\) &
Strong ensemble exposure; selectivity and individuality unresolved \\
Candidate adaptive & High under challenge & Coordinated, task-relevant &
Requires within-context or episode evidence and demonstrated function \\
\hline
\end{tabular}
\end{center}

Algorithmic descriptors add a structural axis to this programme, not an
alternative persistence or viability criterion. Raw BDM must identify its
encoded object; paired response masks require the same-shaped zero-mask
reference, whereas an unperturbed trajectory describes state structure.
Neither a low description length nor a flat curve implies maintained
organisation, and a nonzero zero-referenced contrast does not establish
beneficial regulation.

This programme is compatible with earlier work on open-ended evolution.
Innovation there is defined relative to counterfactual isolated dynamics,
and scalable open-ended evolution was associated with state-dependent
dynamics \cite{adams2017}. The complementary question here is whether an
interacting subsystem preserves an independently specified organisation
while its dynamics change. Neither innovation nor persistence alone
demonstrates biological life.

The directional ratio can also be followed through time,
\[
\alpha(t)=\frac{X(t)-Y(t)}{X(t)+Y(t)},
\qquad X(t)+Y(t)>0.
\]
A hypothesised sensing--integration--action sequence might move from
negative through approximately zero to positive \(\alpha(t)\).
Such switching can also occur in passive or externally driven systems;
its adaptive interpretation requires task-aligned interventions and
functional evidence. Moreover, averaging different contexts at each age
can manufacture an apparent sequence absent from any individual trajectory.
Matched-context time series or explicitly aligned episodes are needed to
test the hypothesis.

Thus persistent organisation, selective reciprocal access and functional
benefit are proposed complementary tests, not consequences of \(C>0\),
\(B>0\), a BDM trend or exact recurrence. The present paper supplies
specified channel measurements, a task-conditional theorem and explicit
counterexamples; the broader classification remains a falsifiable
extension rather than an empirical finding.

\clearpage

\section{Estimation algorithm and verification trail}\label{appendix-a.-estimation-algorithm-and-audit-trail}

For each rule pair and initial configuration, evolve the unperturbed
system to the intervention age and store its reference trajectory
through the required horizon. For every eligible source in \(A\), flip
that source alone, evolve the open system and calculate the eligible
environment response. Run the corresponding environment-clamped arm and
calculate late internal response. Repeat open-system interventions for
the matched environment sources and record the focal response. Average
the source-normalised quantities within the context, then average
primitive quantities across initial conditions. Only then calculate the
aggregate scores.

The primary scan evaluates 32,000 pair-context units. Each unit contains
20 open-system single-source interventions and ten additional clamped
interventions, giving 640,000 open-system and 320,000 clamped
counterfactual trajectories, plus their references. The simulation code
stores mean changed-site counts and geometric opportunities as well as
densities. These counts permit verification of normalisation but are not
interpreted as numbers of independent information bits.

The main experiment seed is 20260905. Statistical and benchmark
calculations use 20260906; independent validation uses 20260907;
graphical interval calculations use 20260908. The recorded numerical
environment is Python 3.13.5, NumPy 2.3.5 and Numba 0.65.1. The source
package records additional library versions. Bootstrap replicates are
5,000 throughout. Code reports rather than conceals ratios that are
undefined because both cross-boundary scores are zero.

A separate pure scalar update function checks the ECA bit-order
convention and evaluates the output entropy of both intervention arms
directly. It agrees with the accelerated damage-indicator estimator in
all tested contexts. Analytic rule checks cover rules 0, 255, 204, 170,
240 and 90. Further tests examine the algebraic bounds, exact
factorisation, Shannon null interventions, boundary cuts, role exchange and
locality-preserving padding. These checks do not supply a numerical BDM
baseline: a separate structural audit must record each encoded object's
shape, estimator settings, raw value and matched zero-array reference.

\section{Exact counterexamples and scope of the causal
claim}\label{appendix-b.-exact-counterexamples-and-scope-of-the-causal-claim}

\textbf{Common input without transmission.} Let a fair bit \(N\) set
both \(A=N\) and \(E=N\), with no causal arrow from \(A\) to \(E\).
Observational \(I(A;E)=1\), but intervening on \(A\) leaves \(E\)
unchanged and the intervention channel has zero information. Correlation
alone therefore fails the causal test.

\textbf{Simple transmission without complexity.} A perfect copy channel
\(R=U\) has one bit of causal information for a fair input despite its
trivial generating rule. High response complexity is not necessary for a
functioning channel.

\textbf{Noise masking a microscopic perturbation.} For \(R=U\oplus N\)
with independent fair future noise, \(I(U;R)=0\) although the two
outputs differ if one artificially fixes \(N\) across intervention arms.
Conditioning on future noise changes the estimand. The ECA damage
identity is exact because those simulations are deterministic, not
because paired Hamming differences estimate every stochastic causal
channel.

\textbf{Joint information invisible to marginal recipients.} For
\(R_1=N\) and \(R_2=N\oplus U\), both \(I(U;R_1)\) and \(I(U;R_2)\)
vanish, but \(I(U;R_1,R_2)=1\). A scalar sum of marginal information
scores cannot replace a joint-channel analysis in general.

\textbf{Stored state without external interaction.} The identity rule
retains a focal bit indefinitely but has no cross-boundary transmission.
Requiring cross-boundary transmission prevents this memory from being
called reciprocal interaction; self-retention alone is not evidence of
self-maintenance.

\textbf{Opposite task meaning under a bijection.} Inverting the response
preserves all channel mutual informations in the regulatory construction
but reverses its relation to the task requirement. Hence sufficiency
cannot be repaired by simply increasing a threshold on \(C\), \(B\) or
\(S\).

\section{Complete example trajectories and additional
results}\label{appendix-c.-complete-example-trajectories-and-additional-results}

The supplementary arrays retain every detailed case at every
intervention age, including the constant, identity, shift, additive and
homogeneous-rule-30 comparisons. The plain-text context table permits
reconstruction of each reported mean without re-running the simulator.
The sensitivity table contains all 126 rule-pair/geometry/horizon
combinations, not only those displayed in the geometry figure in
Appendix~\ref{geometry-and-intervention-controls}.
Exact recurrence outputs include the transient and period of every
observed cycle and an explicit censoring marker otherwise.

\begin{figure}[h]
\centering
\includegraphics[width=4in,height=2.5in]{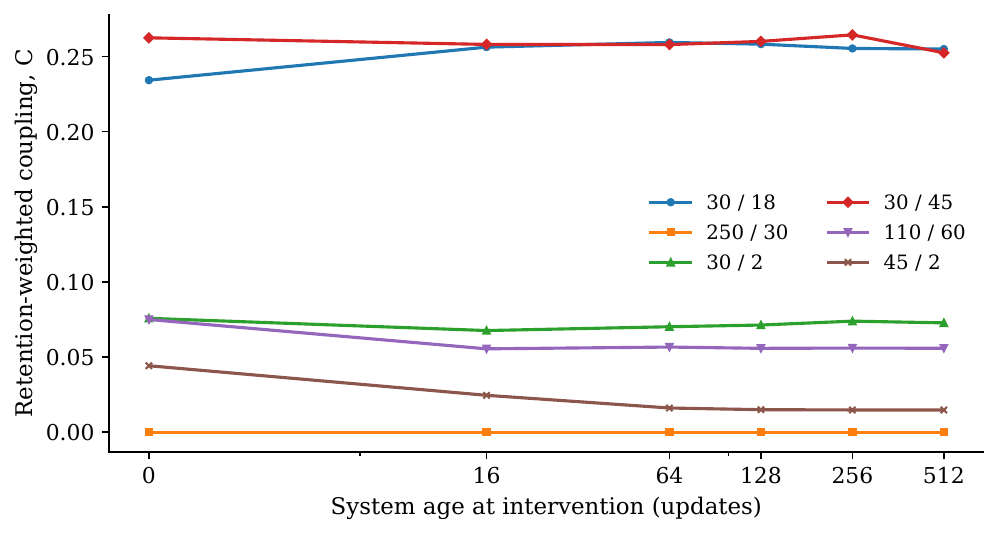}
\caption{Coupling for all six detailed rule pairs across the six
intervention ages. This supplementary view includes the propagation
example and the contracting focal region as well as the three principal
transient comparisons. Numerical outputs and context-level uncertainty
are retained in the data package; the full curves do not establish a
universal monotonic trend.}
\end{figure}

\begin{longtable}[]{@{}
  >{\raggedright\arraybackslash}p{(\columnwidth - 12\tabcolsep) * \real{0.1600}}
  >{\raggedright\arraybackslash}p{(\columnwidth - 12\tabcolsep) * \real{0.1400}}
  >{\raggedright\arraybackslash}p{(\columnwidth - 12\tabcolsep) * \real{0.1400}}
  >{\raggedright\arraybackslash}p{(\columnwidth - 12\tabcolsep) * \real{0.1400}}
  >{\raggedright\arraybackslash}p{(\columnwidth - 12\tabcolsep) * \real{0.1400}}
  >{\raggedright\arraybackslash}p{(\columnwidth - 12\tabcolsep) * \real{0.1400}}
  >{\raggedright\arraybackslash}p{(\columnwidth - 12\tabcolsep) * \real{0.1400}}@{}}
\caption{Homogeneous controls at age 64 under the primary geometry and
horizon. Zero coupling can reflect absent transmission or absent late
internal retention; the primitive components distinguish these
possibilities.}\tabularnewline
\toprule\noalign{}
\begin{minipage}[b]{\linewidth}\raggedright
Rules A / E
\end{minipage} & \begin{minipage}[b]{\linewidth}\raggedright
\(X\)
\end{minipage} & \begin{minipage}[b]{\linewidth}\raggedright
\(Y\)
\end{minipage} & \begin{minipage}[b]{\linewidth}\raggedright
\(S\)
\end{minipage} & \begin{minipage}[b]{\linewidth}\raggedright
\(C\)
\end{minipage} & \begin{minipage}[b]{\linewidth}\raggedright
\(D\)
\end{minipage} & \begin{minipage}[b]{\linewidth}\raggedright
\(\alpha\)
\end{minipage} \\
\midrule\noalign{}
\endfirsthead
\toprule\noalign{}
\begin{minipage}[b]{\linewidth}\raggedright
Rules A / E
\end{minipage} & \begin{minipage}[b]{\linewidth}\raggedright
\(X\)
\end{minipage} & \begin{minipage}[b]{\linewidth}\raggedright
\(Y\)
\end{minipage} & \begin{minipage}[b]{\linewidth}\raggedright
\(S\)
\end{minipage} & \begin{minipage}[b]{\linewidth}\raggedright
\(C\)
\end{minipage} & \begin{minipage}[b]{\linewidth}\raggedright
\(D\)
\end{minipage} & \begin{minipage}[b]{\linewidth}\raggedright
\(\alpha\)
\end{minipage} \\
\midrule\noalign{}
\endhead
\bottomrule\noalign{}
\endlastfoot
0 / 0 & 0 & 0 & 0 & 0 & 0 & Undefined \\
204 / 204 & 0 & 0 & 0.1000 & 0 & 0 & Undefined \\
90 / 90 & 0.2485 & 0.2165 & 0.2638 & 0.1226 & 0.0084 & 0.0688 \\
170 / 170 & 0.0347 & 0.0197 & 0 & 0 & 0 & 0.2768 \\
30 / 30 & 0.3151 & 0.3749 & 0.4035 & 0.2784 & -0.0241 & -0.0866 \\
\end{longtable}


\clearpage
\begin{thebibliography}{99}

\bibitem{zenil2012}
Zenil, H., Gershenson, C., Marshall, J. A. R., and Rosenblueth, D. A.
(2012). Life as thermodynamic evidence of algorithmic structure in
natural environments. \emph{Entropy}, 14, 2173--2191.
doi:\href{https://doi.org/10.3390/e14112173}{10.3390/e14112173}.

\bibitem{krakauer2020}
Krakauer, D., Bertschinger, N., Olbrich, E., Flack, J. C., and Ay, N.
(2020). The information theory of individuality. \emph{Theory in
Biosciences}, 139, 209--223.
doi:\href{https://doi.org/10.1007/s12064-020-00313-7}{10.1007/s12064-020-00313-7}.

\bibitem{kolchinsky2018}
Kolchinsky, A. and Wolpert, D. H. (2018). Semantic information,
autonomous agency and non-equilibrium statistical physics.
\emph{Interface Focus}, 8, 20180041.
doi:\href{https://doi.org/10.1098/rsfs.2018.0041}{10.1098/rsfs.2018.0041}.

\bibitem{schreiber2000}
Schreiber, T. (2000). Measuring information transfer. \emph{Physical
Review Letters}, 85, 461--464.
doi:\href{https://doi.org/10.1103/PhysRevLett.85.461}{10.1103/PhysRevLett.85.461}.

\bibitem{lizier2010}
Lizier, J. T. and Prokopenko, M. (2010). Differentiating information
transfer and causal effect. \emph{The European Physical Journal B}, 73,
605--615.
doi:\href{https://doi.org/10.1140/epjb/e2010-00034-5}{10.1140/epjb/e2010-00034-5}.

\bibitem{ay2008}
Ay, N. and Polani, D. (2008). Information flows in causal networks.
\emph{Advances in Complex Systems}, 11, 17--41.
doi:\href{https://doi.org/10.1142/S0219525908001465}{10.1142/S0219525908001465}.

\bibitem{klyubin2005}
Klyubin, A. S., Polani, D., and Nehaniv, C. L. (2005). All else being
equal be empowered. \emph{Advances in artificial life}, 3630. Springer.
doi:\href{https://doi.org/10.1007/11553090_75}{10.1007/11553090\_75}.

\bibitem{zenil2019}
Zenil, H., Kiani, N. A., Marabita, F., Deng, Y., Elias, S., Schmidt,
A., Ball, G., and Tegn{\'e}r, J. (2019). An algorithmic information calculus
for causal discovery and reprogramming systems. \emph{iScience}, 19,
1160--1172.
doi:\href{https://doi.org/10.1016/j.isci.2019.07.043}{10.1016/j.isci.2019.07.043}.

\bibitem{adamszenilreyesjoosten2015}
Adams, A., Zenil, H., Reyes, E., and Joosten, J.
(2015). The Effects of Global Rules on Interacting Cellular Automata.
In J. Kari, I. T\"orm\"a, and M. Szabados (Eds.),
\emph{21st International Workshop on Cellular Automata and Discrete
Complex Systems (AUTOMATA 2015)}.
TUCS Lecture Notes.

\bibitem{aid2020}
Zenil, H., Kiani, N. A., Abrah{\~a}o, F. S., and Tegn{\'e}r, J. (2020).
Algorithmic information dynamics. \emph{Scholarpedia}, 15, 53143.
doi:\href{https://doi.org/10.4249/scholarpedia.53143}{10.4249/scholarpedia.53143}.

\bibitem{wolfram1984}
Wolfram, S. (1984). Universality and complexity in cellular
automata. \emph{Physica D}, 10, 1--35.
doi:\href{https://doi.org/10.1016/0167-2789(84)90245-8}{10.1016/0167-2789(84)90245-8}.

\bibitem{derrida1986}
Derrida, B. and Pomeau, Y. (1986). Random networks of automata: A
simple annealed approximation. \emph{Europhysics Letters}, 1, 45--49.
doi:\href{https://doi.org/10.1209/0295-5075/1/2/001}{10.1209/0295-5075/1/2/001}.

\bibitem{cover2006}
Cover, T. M. and Thomas, J. A. (2006). Elements of information
theory. Wiley.

\bibitem{bdm2018}
Zenil, H., Hern\'andez-O, A., and Tegn\'er, J. (2018). A decomposition method for. A decomposition method for
global evaluation of Shannon Entropy and local estimations of
Algorithmic Complexity. \emph{Entropy}, 20, 605.
doi:\href{https://doi.org/10.3390/e20080605}{10.3390/e20080605}.

\bibitem{uthamacumaran2024}
Uthamacumaran, A., Abrah{\~a}o, F. S., Kiani, N. A., and Zenil, H.
(2024). On the salient limitations of the methods of assembly theory and
their classification of molecular biosignatures.
\emph{npj Systems Biology and Applications}, 10, 82.
doi:10.1038/s41540-024-00403-y.

\bibitem{abrahao2024}
Abrah{\~a}o, F. S., Hern{\'a}ndez-Orozco, S., Kiani, N. A.,
Tegn{\'e}r, J., and Zenil, H.
(2024). Assembly Theory is an approximation to algorithmic complexity
based on LZ compression that does not explain selection or evolution.
\emph{PLOS Complex Systems}, 1(1), e0000014.
doi:10.1371/journal.pcsy.0000014.

\bibitem{ozelim2026}
Ozelim, L., Uthamacumaran, A., Abrah{\~a}o, F. S.,
Hern{\'a}ndez-Orozco, S., Kiani, N. A., Tegn{\'e}r, J.,
Hern{\'a}ndez-Quiroz, F., and Zenil, H.
(2026). Assembly theory collapses to dictionary compression and is
rendered redundant by common statistical algorithms.
\emph{npj Complexity}.
doi:10.1038/s44260-026-00088-w.

\bibitem{barricelli1954}
Barricelli, N. A. (1954). Esempi numerici di processi di evoluzione.
\emph{Methodos}, 6, 45--68.

\bibitem{bialek2021}
Bialek, W. and Schwab, D. J. (2021). The physics of life.
\emph{Annual Review of Condensed Matter Physics}, 12, 1--24.

\bibitem{gardner1970}
Gardner, M. (1970). Mathematical games: The fantastic combinations of
John Conway's new solitaire game ``Life''.
\emph{Scientific American}, 223(4), 120--123.

\bibitem{eigen1971}
Eigen, M. (1971). Selforganisation of matter and the evolution of
biological macromolecules. \emph{Naturwissenschaften}, 58, 465--523.

\bibitem{england2013}
England, J. L. (2013). Statistical physics of self-replication.
\emph{Journal of Chemical Physics}, 139, 121923.

\bibitem{langton1989}
Langton, C. G. (1989). Artificial life. In \emph{Artificial Life I},
pp. 1--47. Addison-Wesley.

\bibitem{livitanyi2019}
Li, M. and Vit\'anyi, P. (2019). \emph{An Introduction to Kolmogorov
Complexity and Its Applications}, 4th ed. Springer.

\bibitem{parrondo2015}
Parrondo, J. M. R., Horowitz, J. M., and Sagawa, T. (2015).
Thermodynamics of information. \emph{Nature Physics}, 11(2), 131--139.

\bibitem{prigogine1980}
Prigogine, I. (1980). \emph{From Being to Becoming: Time and Complexity
in the Physical Sciences}. W. H. Freeman.

\bibitem{schrodinger1944}
Schr\"odinger, E. (1944). \emph{What Is Life?} Cambridge University Press.

\bibitem{shannon1948}
Shannon, C. E. (1948). A mathematical theory of communication.
\emph{Bell System Technical Journal}, 27, 379--423 and 623--656.

\bibitem{solomonoff1964}
Solomonoff, R. J. (1964). A formal theory of inductive inference, Parts I
and II. \emph{Information and Control}, 7, 1--22 and 224--254.

\bibitem{wissnergross2013}
Wissner-Gross, A. D. and Freer, C. E. (2013). Causal entropic forces.
\emph{Physical Review Letters}, 110, 168702.

\bibitem{ctm2014}
Soler-Toscano, F., Zenil, H., Delahaye, J.-P., and Gauvrit, N. (2014).
Calculating Kolmogorov complexity from the output frequency distributions
of small Turing machines. \emph{PLOS ONE}, 9(5), e96223.
doi:10.1371/journal.pone.0096223.

\bibitem{zenil2007compressibility}
Zenil, H. (2007).
Cellular Automaton Compressibility.
\emph{Wolfram Demonstrations Project}.
Wolfram Research.

\bibitem{zenilvillarreal2014}
Zenil, H. and Villarreal, E. (2014).
Cellular Automata Sensitivity to Initial Conditions.
\emph{Wolfram Demonstrations Project}.
Wolfram Research.

\bibitem{zenilmarshall2013}
Zenil, H. and Marshall, J. A. R. (2013). Some Aspects of Computation
Essential to Evolution and Life. \emph{ACM Ubiquity}, 2013(April), 1--16.
doi:10.1145/2480352.2480353.

\bibitem{adams2017}
Adams, A., Zenil, H., Davies, P. C. W., and Walker, S. I. (2017).
Formal Definitions of Unbounded Evolution and Innovation Reveal Universal
Mechanisms for Open-Ended Evolution in Dynamical Systems.
\emph{Scientific Reports}, 7, 997.
doi:10.1038/s41598-017-00810-8.

\bibitem{hernandez2018}
Hern{\'a}ndez-Orozco, S., Hern{\'a}ndez-Quiroz, F., and Zenil, H. (2018).
The Limits of Decidable States on Open-Ended Evolution and Emergence.
\emph{Artificial Life}, 24(1), 56--70.

\bibitem{zenil2015gol}
Zenil, H. (2015). Algorithmicity and Programmability in Natural
Computing with the Game of Life as an In Silico Case Study.
\emph{Journal of Experimental \& Theoretical Artificial Intelligence},
27(1), 109--121.
doi:10.1080/0952813X.2014.940686.

\bibitem{zenil2019gol}
Zenil, H., Kiani, N. A., and Tegn{\'e}r, J. (2019). Algorithmic
Information Dynamics of Emergent, Persistent, and Colliding Particles in
the Game of Life. In \emph{From Parallel to Emergent Computing},
pp. 367--384. CRC Press.
doi:10.1201/9781315167084-17.

\bibitem{zeniladams2022}
Zenil, H. and Adams, A. (2022). Algorithmic Information Dynamics of
Cellular Automata. In \emph{The Mathematical Artist: A Tribute to John
Horton Conway}, pp. 171--190. Springer.
doi:10.1007/978-3-031-03986-7\_8.

\bibitem{abrahao2022observers}
Abrah{\~a}o, F. S. and Zenil, H. (2022). Emergence and Algorithmic
Information Dynamics of Systems and Observers.
\emph{Philosophical Transactions of the Royal Society A}, 380(2227),
20200429.
doi:10.1098/rsta.2020.0429.

\bibitem{gauvritzenil2017}
Gauvrit, N., Zenil, H., and Tegn{\'e}r, J. (2017). The
Information-Theoretic and Algorithmic Approach to Human, Animal, and
Artificial Cognition. In \emph{Representation and Reality in Humans,
Other Living Organisms and Intelligent Machines}, pp. 117--139. Springer.
doi:10.1007/978-3-319-43784-2\_7.

\bibitem{zenilcognition2023}
Zenil, H., Marshall, J. A. R., and Tegn{\'e}r, J. (2023).
Approximations of Algorithmic and Structural Complexity Validate
Cognitive-Behavioral Experimental Results.
\emph{Frontiers in Computational Neuroscience}, 16, 956074.
doi:10.3389/fncom.2022.956074.

\bibitem{ruffini2017}
Ruffini, G. (2017). An Algorithmic Information Theory of Consciousness.
\emph{Neuroscience of Consciousness}, 2017(1), nix019.
doi:10.1093/nc/nix019.

\bibitem{ruffini2026regulator}
Ruffini, G. (2026). The Algorithmic Regulator.
\emph{Entropy}, 28(3), 257.
doi:10.3390/e28030257.

\bibitem{ruffini2026patterns}
Ruffini, G. and Klaus (2026). When Patterns Persist: Reach, Regulation,
and Death in the Game of Life. \emph{BCOM Working Paper WP0191},
companion to WP0162. Zenodo.
doi:10.5281/zenodo.21008860.

\bibitem{horowitz2014}
Horowitz, J. M. and Esposito, M. (2014). Thermodynamics with
continuous information flow. \emph{Physical Review X}, 4, 031015.
doi:\href{https://doi.org/10.1103/PhysRevX.4.031015}{10.1103/PhysRevX.4.031015}.

\end{thebibliography}
\end{document}